\documentclass[
 preprints,article,
 accept,pdftex,oneauthor]{Definitions/mdpi} 
\firstpage{1} 
\pubvolume{1}
\issuenum{1}
\articlenumber{0}
\pubyear{2022}
\copyrightyear{2022}
\datereceived{} 
\dateaccepted{} 
\datepublished{} 
\hreflink{https://doi.org/} 

\Title{CP Violation for the Heavens and the Earth}

\TitleCitation{CPV for BAU and $e$EDM}

\Author{George Wei-Shu Hou $^{1
}$\orcidA{}}

\AuthorNames{George Wei-Shu Hou}

\AuthorCitation{Hou, G.W.S.}

\address{%
$^{1}$ \quad Department of Physics, National Taiwan University, Taipei 10617, Taiwan; wshou@phys.ntu.edu.tw
}

\abstract{
Electroweak baryogenesis can be driven by 
the top quark in a general two Higgs doublet model 
with extra Yukawa couplings. Higgs quartics provide 
the first order phase transition, while 
extra top Yukawa coupling $\rho_{tt}$ 
can fuel the cosmic baryon asymmetry through 
the $\lambda_t\,{\rm Im}\,\rho_{tt}$ product, with 
flavor-changing $\rho_{tc}$ coupling as backup. 
The impressive ACME 2018 bound on the electron 
electric dipole moment calls for an extra 
electron coupling $\rho_{ee}$ for exquisite 
cancellation among dangerous diagrams, 
broadening the baryogenesis solution space. 
The mechanism suggests that extra Yukawa couplings echo 
the hierarchical structure of standard Yukawa couplings. 
Phenomenological consequences in Higgs search 
and flavor physics are discussed, 
with $\mu$ and $\tau$ EDM touched upon.
}


\begin{document}


\section{Introduction: Our Life and Times}

The 125 GeV boson, $h$, was discovered in 2012, 
but No New Physics (NNP) beyond the Standard Model
 (BSM) has been found: 
 not before 2012, and not in the decade since. 
Where is SUSY \cite{ref-proceeding1}, 
 the long-anticipated front-runner? 
 And where is everybody else?

One place where people converge on are ``Out-of-the-Box'' searches, from \cite{ref-journal1} 
ALPs (Axion-Like Particles) to \cite{ref-journal2} 
LLPs (Long-Lived Particles), reflecting 
a sign of our times, as if we built the grand 
``boxes'' called ATLAS and CMS, etc. out of human vanity. 
But really? 
It also reflects the lack of evidence for
 \cite{ref-journal3} WIMPs 
 (Weakly Interacting Massive Particles) 
 after a plethora of Dark Matter searches,
 where SUSY had provided excellent candidates. 

Another general direction that is in vogue is 
EFT \cite{ref-proceeding2}, or Effective Field Theory: 
since No~New~Particles (NNP) are seen other than 
those of SM, one assumes New states exist above 
some ``cutoff'' scale $\Lambda$, far above 
the known SM particles such as $t$, $h$, $Z/W$ 
that are below the v.e.v. scale of 246 GeV. 
The latter give the dimension-4 terms of 
the SM Lagrangian, while we can only (nominally) 
divine minute deviations from SM with dimension-6 
or higher operators as an expansion in $1/\Lambda$.

But have we truly exhausted ``normal'' dimension-4 terms?
We wish to explore something 
``unconventionally-conventional'', 
a Road Not Taken (by most): we advocate 
the existence of an {\it extra} Higgs doublet that 
carries {\it extra} Yukawa couplings, and of course 
the accompanying {\it extra} Higgs quartic couplings. 
We argue that the exotic Higgs bosons, $H$, $A$ and $H^+$, 
are naturally sub-TeV in mass. Hence, the two sets of 
new dimension-4 couplings should be pursued at the LHC,
''within the box'' of ATLAS and CMS, 
as well as at the flavor frontier.

An {\it extra} Higgs doublet sounds conventional enough, 
but in part influenced by SUSY, 
an {\it extra} Higgs doublet is usually viewed as 
without {\it extra}, or a second set of Yukawa couplings. 
We will show how this is very much a prejudice, 
as our perspective is broadened by considering 
other big issues, such as baryogenesis, which 
calls for additional, {\it large} CP violation
 (CPV) sources. Since all experimentally verified 
  CPV so far \cite{ref-journal4} 
come from Yukawa couplings, we place special premium 
on these {\it extra} Yukawa couplings. 
Furthermore, electroweak baryogenesis (EWBG) comes 
closer to heart as it is more testable, 
perhaps even at the LHC. 
This makes a second Higgs doublet with 
extra Yukawa couplings attractive, since 
as a cousin of the observed Higgs boson doublet, 
we should explore all its possible aspects at 
the electroweak scale. But NNP at the LHC poses 
a challenge: Can one still have large CPV for EWBG? 
We will show that the answer is quite in the affirmative. 
What is more, the {\it extra} Higgs quartics gets 
thrown in as a bonus to provide first order 
electroweak phase transition (EWPT), 
one of the three Sakharov conditions
 \cite{ref-journal5} for baryogenesis.

Billions and billions of stars, all those protons burning 
to light the Universe --- but no antiprotons! 
The Baryon Asymmetry of the Universe (BAU), or 
disappearance of antimatter very shortly 
after the Big Bang, is indeed a problem 
as big as the Universe itself, and at 
the very core of our own existence.

CPV for the Heavens, or having {\it extra}, 
large sources of CPV for EWBG sounds attractive, 
but this brings about another problem: 
precision low energy (LE) experiments {\it on Earth}, 
such as ACME, which has recently pushed the bound on 
{\it electron} electric dipole moment ($e$EDM) down to 
the very impressive $10^{-29}\; e$\,cm level 
 \cite{ref-journal6}. Can an {\it extra, large} CPV source
 that drives EWBG survive such stringent 
 precision frontier LE probes? Indeed, table-top 
 experiments like ACME are now competing directly 
 with behemoths such as the LHC and its associated 
 experiments at the high energy (HE) frontier. 

Perhaps a bit surprisingly, we will show that {\it extra} 
Yukawa couplings of an {\it extra} Higgs doublet may 
come with the finesse to survive ACME's check on EWBG 
--- CPV on Earth! The finesse echoes the mysterious 
``flavor enigma'': the {\it mass hierarchy} 
between fermion generations, and the apparently 
correlated {\it mixing hierarchy} of the three 
mixing angles of the Cabibbo-Kobayashi-Maskawa 
(CKM) matrix \cite{ref-journal4}, $V$. 
{\it Nature} even threw in the recently emerged ``alignment'' phenomenon, that the observed 
$h$ boson does not seem to mix (much) with 
the exotic CP-even $H$ boson.
While unrelated to flavor per se, it helps hide 
this {\it extra} Higgs doublet from our view, so far. 
All these we shall elucidate. 

In the following, we first present the {\it general} 
two Higgs doublet model (g2HDM) and illustrate how 
it can bring about EWBG. We then show how g2HDM 
pulls the finesse to survive the ACME bound, which 
{\it implies $e$EDM could be just around the corner.} 
On the wheres and hows to verify this BSM physics,
important phenomenological consequences are then 
discussed. 
Foretelling our Summary, the g2HDM is really just 
SM but with two Higgs doublets (SM2). 
It is thus quite simple, but dynamical parameters abound, 
providing {\bf CPV for the Heavens and the Earth}
--- where {\it general} 2HDM offers an illustration.

 %
 %

\section{General Two Higgs Doublet Model}

Out of the three Sakharov conditions \cite{ref-journal5} 
for baryogenesis, his original suggestion of
baryon number violation is provided by
electroweak theory at high temperature, 
which is realized at the very early Universe.
But a first order electroweak phase transition (EWPT)?
Or {\it sufficient} amount of CPV?
On these two counts, SM falls short:
the weak interaction, as well as 
the SM Higgs quartic coupling $\lambda$, are too weak;
the Jarlskog invariant of SM falls way too short
 \cite{ref-journal7}, receiving high powers of 
light mass as well as CKM suppression.

Adding a second Higgs doublet, i.e. 2HDM, can help:
The first order EWPT is achievable with
${\cal O}(1)$ Higgs quartics in the Higgs potential, 
$V(\Phi_1, \Phi_2)$. Conventional wisdom is that 
one may want to keep the Higgs potential CP conserving, 
otherwise one could run into problems with 
electric dipole moments, such as $d_n$ of the neutron.

As we have mentioned, all measured CPV in the laboratory
can be accounted for by the CKM matrix $V$,
which originates from Yukawa couplings.
Thus, for sake of baryogenesis, extra BSM Yukawa couplings, 
such as due to a second Higgs doublet, should be welcome.
However, having two Yukawa coupling matrices, 
say for up-type quarks, the linear combination 
that is orthogonal to the mass matrix {\boldmath $m$}$^u$
cannot be simultaneously diagonalized.
This means the presence of flavor changing neutral Higgs 
(FCNH) couplings, the fear of which led
Glashow and Weinberg to famously propose \cite{ref-journal8} 
the Natural Flavor Conservation (NFC) condition,
that each type of charged fermion receive mass from 
just one scalar doublet, thereby killing FCNH couplings.
The NFC condition is usually implemented in 2HDM
via a $Z_2$ symmetry, leading to 2HDM type I and type II,
where the latter is automatic in minimal SUSY 
for separate reasons, making it rather popular.

But if one puts a premium on extra Yukawa couplings
for sake of baryogenesis, imposing a $Z_2$ symmetry 
would seem {\it ad hoc} and an overkill.
After all, with NNP seen at the LHC, 
any add-on symmetry should be suspect.
We therefore wish to explore the 2HDM further
{\it without} any $Z_2$ symmetry. 
Without a $Z_2$ to impose NFC, 
we have a second set of Yukawa couplings, 
e.g. {\boldmath $\rho$}$^u$ to accompany the 
SM Yukawa matrix {\boldmath 
$\lambda$}$^u \equiv \sqrt 2${\boldmath $m$}$^u/v$, 
where $v \simeq 246$\,GeV is the vacuum expectation value. 
We call this the {\it general} 2HDM, or g2HDM.

In the next subsection, we will show that the up-type
extra Yukawa coupling $\rho_{tt}$, expected 
at ${\cal O}(1)$ by analogy with $\lambda_t \cong 1$,
can drive EWBG, while $\rho_{tc}$ provides a backup.
So let us prepare for the two needed elements for EWBG:
Extra Yukawa couplings as CPV source, and
extra Higgs quartics to provide first order EWPT.
We will address the concerns of Glashow and Weinberg later.

\subsection{Extra Yukawa Couplings}

With two Higgs doublets and without a $Z_2$ coupling 
to enforce NFC, the general Yukawa couplings for 
up-type quarks are
\begin{equation}
   \bar u_{iL}(Y^u_{1ij} \tilde \Phi_1
             + Y^u_{2ij} \tilde \Phi_2) u_{jR} 
             + {\rm h.c.}
\label{eq:Yukawa}
\end{equation}
Although later on we would invoke the Higgs basis and
put the v.e.v. to only one Higgs doublet,
as the Universe cooled down from its very hot beginning
and electroweak symmetry breaking develops,
the evolution history can be viewed as effectively 
tracing through both doublets having v.e.v.s. Thus, 
we will take $v_1 = v\cos\beta$ and $v_2 = v\sin\beta$
where $\beta$ is temperature-dependent, and 
\begin{equation}
    Y^{u}_{\rm SM} = Y_1^u \cos\beta + Y_2^u \sin\beta,
 \label{eq:Y_uSM}
\end{equation}
feeds the mass matrix, which is
diagonalized by $U_L^{\dagger} Y^{u}_{\rm SM} U_R
 = {\rm diag}(\lambda_u, \lambda_c, \lambda_t)$ as usual.
The orthogonal combination,
\begin{equation}
    \rho^{u} = U_L^{\dagger}
              (- Y_1^u \sin\beta + Y_2^u \cos\beta)
               U_R,
 \label{eq:rho_u}
\end{equation}
then cannot be simultaneously diagonal, 
leading to FCNH couplings.

The $\rho^u$ matrix is orthogonal to the mass matrix.
But the Higgs potential would induce mixing
between the light $h$ boson from the mass giving doublet 
and the exotic CP-even boson $H$, 
and we arrive at the neutral up-type Yukawa 
interaction at $T = 0$:
\begin{align}
   \bar u_{iL} \left(
    \frac{\lambda_i \delta_{ij}}{\sqrt2} s_{\gamma}
  + \frac{\rho_{ij}^u}{\sqrt2} c_{\gamma}
               \right) u_{jR}\, h 
  +\bar u_{iL} \left(
    \frac{\lambda_i \delta_{ij}}{\sqrt2} c_{\gamma}
  - \frac{\rho_{ij}^u}{\sqrt2} s_{\gamma}
               \right) u_{jR}\, H 
  -\frac{i}{\sqrt2}\bar u_{iL}^u\rho_{ij}u_{jR}\, A
  + {\rm h.c.},
 \label{eq:Yuk_u}
\end{align}
where $c_\gamma = \cos(\beta - \alpha)$ is the
$h$-$H$ mixing angle in SUSY notation.
The recently emergent ``alignment'' phenomenon
states that, if a second Higgs doublet exists,
$h$-$H$ mixing seems small \cite{ref-journal9}.
We see that, in the alignment limit of $c_\gamma \to 0$,
the $h$ couplings become diagonal, even if 
the $\rho$ matrix is nondiagonal.
This helps control processes such as $t \to ch$ decay,
where CMS set recently \cite{ref-unpublish1} 
the most stringent bound.
If the effect of the extra $\rho$ Yukawa matrix
on $h$ disappears in the alignment limit,
its full effect is in the exotic scalar sector.

The fact that LHC experiments search for
$t \to ch$ and $h \to \tau\mu$ \cite{ref-journal4} 
states that, whether NFC is active or not in 
{\it Nature} is actually an experimental question.
Note also from Eq.~(\ref{eq:Yuk_u}) that,
due to the chiral nature of the weak interactions,
the extra Yukawa couplings, as in SM, are complex,
i.e. $\rho_{ij} = |\rho_{ij}|e^{i\phi_{ij}}$,
which we employ towards baryogenesis.

\subsection{Extra Higgs Quartic Couplings}

Besides source of CPV, the other prerequisite 
for EWBG is to have a first order EWPT, which SM lacks,
but 2HDM can provide, {\it if} extra Higgs quartics
are ${\cal O}(1)$ \cite{ref-journal10}.

The most general CP-conserving potential of g2HDM
in the Higgs basis is \cite{ref-journal11, ref-journal12}
\begin{align}
 V(\Phi,\,\Phi')
&=\      \mu_{11}^2 |\Phi|^2 +\mu_{22}^2 |\Phi'|^2
 - \left(\mu_{12}^2 \Phi^\dagger\Phi' +{\rm h.c.}\right) 
 +\frac{\eta_1}{2}|\Phi|^4 + \frac{\eta_2}{2}|\Phi'|^4
  \notag\\
& + \eta_3|\Phi^2||\Phi'|^2 + \eta_4|\Phi^\dagger\Phi'|^2 
   +\left\{\left(\frac{\eta_5^{}}{2}\Phi^\dagger\Phi'
    +\eta_6^{}|\Phi|^2 + \eta_7^{}|\Phi'|^2\right)
        \Phi^\dagger\Phi' + {\rm h.c.}\right\},
\label{eq:pote2}
\end{align}
where $\mu_{11}^2 < 0$ generates v.e.v., $v \neq 0$,
with a slightly different convention from SM potential.
A second minimization condition,
 $\mu_{12}^2 = \eta_6 v^2/2$,
eliminates $\mu_{12}^2$ as a parameter, 
leaving $\eta_6$ as the sole parameter for 
$h$-$H$ mixing, i.e. $c_\gamma$.
Note that $\eta_6$ and $\eta_7$ terms 
would be absent with usual $Z_2$ symmetry.
We note that the potential of Eq.~(\ref{eq:pote2})
makes better sense than the usual ones under $Z_2$:
by putting symmetry breaking in $\mu_{11}^2 < 0$,
we have $\mu_{22}^2 > 0$ as inertial mass
while $\mu_{12}^2$ is eliminated.
In contrast, in 2HDM I \& II, one has
both $\mu_{11}^2 < 0$ and $\mu_{22}^2 < 0$,
while $\mu_{12}^2$ serves the dual function 
as inertial mass and $h$-$H$ mixing parameter.

One sees now that requiring a first order EWPT, i.e. having 
extra Higgs quartics at ${\cal O}(1)$ \cite{ref-journal10}
has implications on $H$, $A$, $H^+$ masses.
One can also argue for $\mu_{22}^2/v^2 = {\cal O}(1)$,
for if it is much larger, it would damp away 
all amplitudes for baryogenesis.
One therefore finds \cite{ref-journal12} that 
the exotic scalars are sub-TeV in mass.
It is interesting to note that,
having all $\eta_i$'s and $\mu_{22}^2/v^2$ 
at ${\cal O}(1)$ reflects ``common'' naturalness
that one learned in high school, e.g.
one does not put one dollar and 1 million dollars
on the table at the same time, whatever the currency.
In any case, the sub-TeV spectrum
 {\it should be fully explored at the LHC} 
before one heads for heavier masses.

It is worthy of note that, one has the 
approximate relation \cite{ref-journal12} near alignment,
\begin{align}
c_\gamma \cong \frac{\eta_6 v^2}{m_H^2 - m_h^2},
 \label{eq:c_gam}
\end{align}
since $s_\gamma$ approaches 1 faster 
than $c_\gamma$ approaches 0.
With $m_h$ fixed at 125 GeV,
one sees that for $m_H \lesssim 300$ GeV or so,
small $c_\gamma$ can be sustained only by 
tuning $\eta_6$ towards zero. So, 
from the ``common naturalness'' perspective, $m_H$ or 
the exotic scalar masses below $v$ is not favored. 
Thus, the target mass range for $H$, $A$, $H^+$
is approximately (300,\, 600) GeV.
But LHC should of course leave no stone unturned.

\section{The Heavens: ElectroWeak BaryoGenesis}

It has been shown that a strongly first order EWPT
can be achieved \cite{ref-journal10} by extra thermal loops 
in 2HDM with ${\cal O}(1)$ Higgs quartic couplings,
fulfilling one of the prerequisites for EWBG.
This we shall assume, taking in fact $H$, $A$ 
and $H^+$ to be degenerate at 500~GeV to simplify.
The main purpose of this section is to illustrate
the CPV source \cite{ref-journal13} in g2HDM.

Let us give an account of EWBG at the semi-folklore level.
Shortly after the Big Bang, one has an expanding bubble 
of the broken phase. Inside the bubble where $v \neq 0$,
baryon number is conserved, but outside the bubble in
the symmetric phase, baryon number is violated by sphalerons.
To avoid baryon number $n_B$ washout, one needs
$\Gamma_B^{(\text{br})}(T_C)<H(T_C)$ in the broken phase,
i.e. the $n_B$ changing rate $\Gamma_B^{(\text{br})}(T_C)$ 
is less than the Hubble parameter $H(T_C)$ 
at critical temperature $T_C$.
To satisfy this condition, first order EWPT 
is needed to ensure \cite{ref-journal14} 
$v_c/T_C > \zeta_{\rm sph}(T_C) = {\cal O}(1)$, where 
$v_C = \sqrt{v_1^2(T_C)+v_2^2(T_C)}$ is the v.e.v. at $T_C$.

The task then is to estimate BAU, i.e. the ratio of
$n_B$ to entropy density $s$, by
\begin{align}
Y_B \equiv \frac{n_B}{s} = \frac{-3\Gamma_B^{(\text{sym})}}{2D_q\lambda_+s}
\int_{-\infty}^0dz'~n_L(z')e^{-\lambda_-z'},
\label{eq:YB}
\end{align}
where $\Gamma_B^{(\text{sym})} = 120\alpha_W^5T$ is 
the $B$ changing rate in symmetric phase,
$D_q \simeq 8.9/T$ is the quark diffusion constant, 
and $\lambda_\pm \simeq v_w$ is the bubble wall velocity,
with $\alpha_W$ the weak coupling constant and $v_w$ 
the bubble wall velocity. One integrates over $z'$, 
the coordinate opposite the bubble expansion direction,
and collect left-handed fermion number 
density $n_L$ inside the bubble as it expands
to become our Universe.
The observed BAU by Planck 2014 is \cite{ref-journal15}
$Y_B^{\rm obs} = 8.59\times 10^{-11}$.

\begin{figure}[t]
\center
\includegraphics[width=3.5cm]{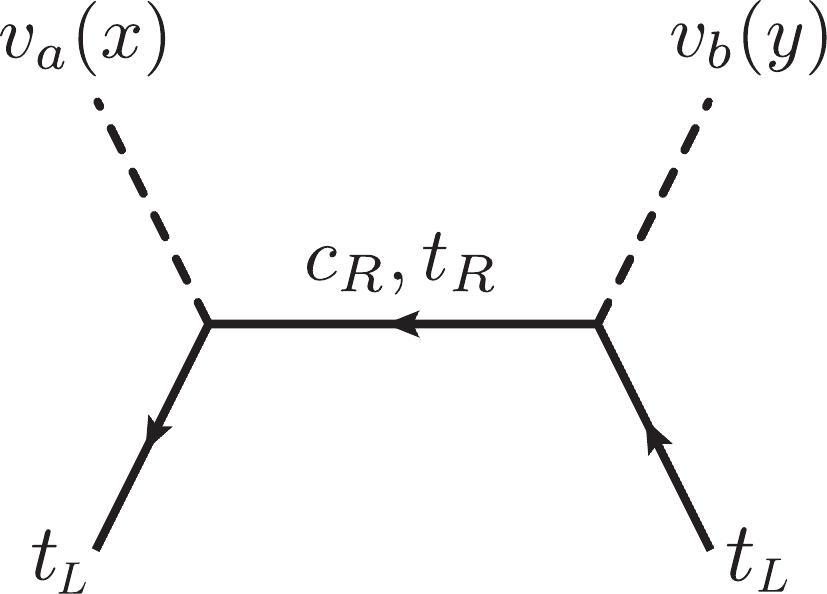}
\caption{
CPV process for generating left-handed top density 
$n_L$ towards BAU, where Higgs bubble wall is denoted 
symbolically as $v_{a}(x)$ and $v_b(y)$.
}
\label{fig:bubble_int}
\end{figure}

In g2HDM, $n_L$ is likely the left-handed top density
via CPV top interactions at the bubble wall,
as illustrated in Fig.~\ref{fig:bubble_int},
where vertices are given in Eq.~(\ref{eq:Yukawa}).
The bubble wall is denoted as the 
spacetime-dependent v.e.v.s \cite{ref-journal14},
$v_{a}(x)$, $v_{b}(y)$ ($a,\,b=1,\,2$).

We skip the discussion of transport equations, which 
we refer to Refs. \cite{ref-journal13, ref-journal14}.

\subsection{CPV Top Interactions}

Skipping details, the CPV source term 
$S_{ij}$~\cite{ref-journal13}
for left-handed fermion $f_{iL}$ induced 
by right-handed fermion $f_{jR}$ is,
\begin{align}
{S_{i_L j_R}(Z)}&= N_C F\,
    \text{Im}\big[(Y_1)_{ij}(Y_2)_{ij}^*\big]\, v^2(Z)\, \partial_{t_Z}\beta(Z),
\label{eq:sourCPV}
\end{align}
where $Z = (t_Z,0,0,z)$ is the position 
in heat bath of the very early Universe,
$N_C = 3$ the color factor, 
and $F$ is a function~\cite{ref-journal14} of 
complex energies of $f_{iL}$ and $f_{jR}$ that 
incorporate $T$-dependent widths of particle/hole modes.
We note that, although the angle $\beta$ is
basis-dependent in g2HDM, its variation
$\partial_{t_Z}\beta(Z)$, reflecting 
the departure from equilibrium, is physical.
We use~\cite{ref-journal13} the value 
$\Delta \beta = 0.015$.

The essence of the CPV for BAU is clearly in $\text{Im}\big[(Y_1)_{ij}(Y_2)_{ij}^*\big]$.
From Eqs.~(\ref{eq:Y_uSM}) and (\ref{eq:rho_u})
and reversing the diagonalization, one has
\begin{align}
\text{Im}\big[(Y_1)_{ij}(Y_2)_{ij}^*\big]
= \text{Im}\big[(U_LY_{\rm diag} U_R^{\dagger})_{ij}
                (U_L\rho U_R^{\dagger})_{ij}^* \big].
\label{eq:YCPV}
\end{align}
A simple exercise \cite{ref-journal16} can help 
one gain understanding, and reflect what may be truly 
happening for the up-type extra Yukawa matrix $\rho$.
Suppose one picks 
$(Y_1)_{tc} \neq 0$, $(Y_2)_{tc} \neq 0$, and
$(Y_1)_{tt} = (Y_2)_{tt} \neq 0$ and all else vanish, 
i.e. altogether 3 (complex) parameters.
Setting $\tan\beta = 1$ for convenience, 
one can easily show that $\sqrt 2 Y^{\rm SM} = Y_1 + Y_2$ 
is diagonalized by just $U_R$ to a single nonvanishing 
33 element $\lambda_t$, the observed SM Yukawa coupling, 
while the combination $-Y_1 + Y_2$ is not diagonalized.
Solving for $U_R$ in terms of nonvanishing elements
in $Y_1$ and $Y_2$, one finds \cite{ref-journal13}
(since $\lambda_t$ is real)
\begin{align}
  \text{Im}\big[ (Y_1)_{tc}(Y_2)_{tc}^* \big]
 = - \lambda_t\, \text{Im}\,\rho_{tt},
\label{eq:simpCPV}
\end{align}
with $\rho_{ct} = 0$, which is part of the construction,
as $\rho_{ct}$ is severely constrained 
by $B_q$--$\bar B_q$ mixings \cite{ref-journal17}.
The less constrained $\rho_{tc}$, 
though related to $\rho_{tt}$, remains a free parameter.

Eq.~(\ref{eq:simpCPV}) is quite remarkable.
With $\lambda_t \cong 1$ affirmed recently by experiment
 \cite{ref-journal4}, the best guess for
$\rho_{tt}$, hence $\text{Im}\,\rho_{tt}$, 
is also ${\cal O}(1)$.
Thus, the CPV source of Eq.~(\ref{eq:simpCPV}) 
is ${\cal O}(1)$ in strength, which is in strong contrast 
to the rather suppressed Jarlskog invariant of 
SM \cite{ref-journal7}, with the SM and extra 
top Yukawa couplings joining forces together.

In a similar vein of having extra Yukawa couplings,
we previously advocated \cite{ref-journal7} 
the fourth generation (4G) as driver of EWBG. 
However, not only one did not find 4G while 
it offered no handle on the order of EWPT,
the dominant Jarlskog invariant (Appendix A) still suffers 
$m_b^2/v^2$ as well as CKM suppression \cite{ref-journal7}.
Providing both the CPV source and first order EWPT, 
the g2HDM is a winner.

\begin{figure}[t]
\center
\includegraphics[width=6.7cm]{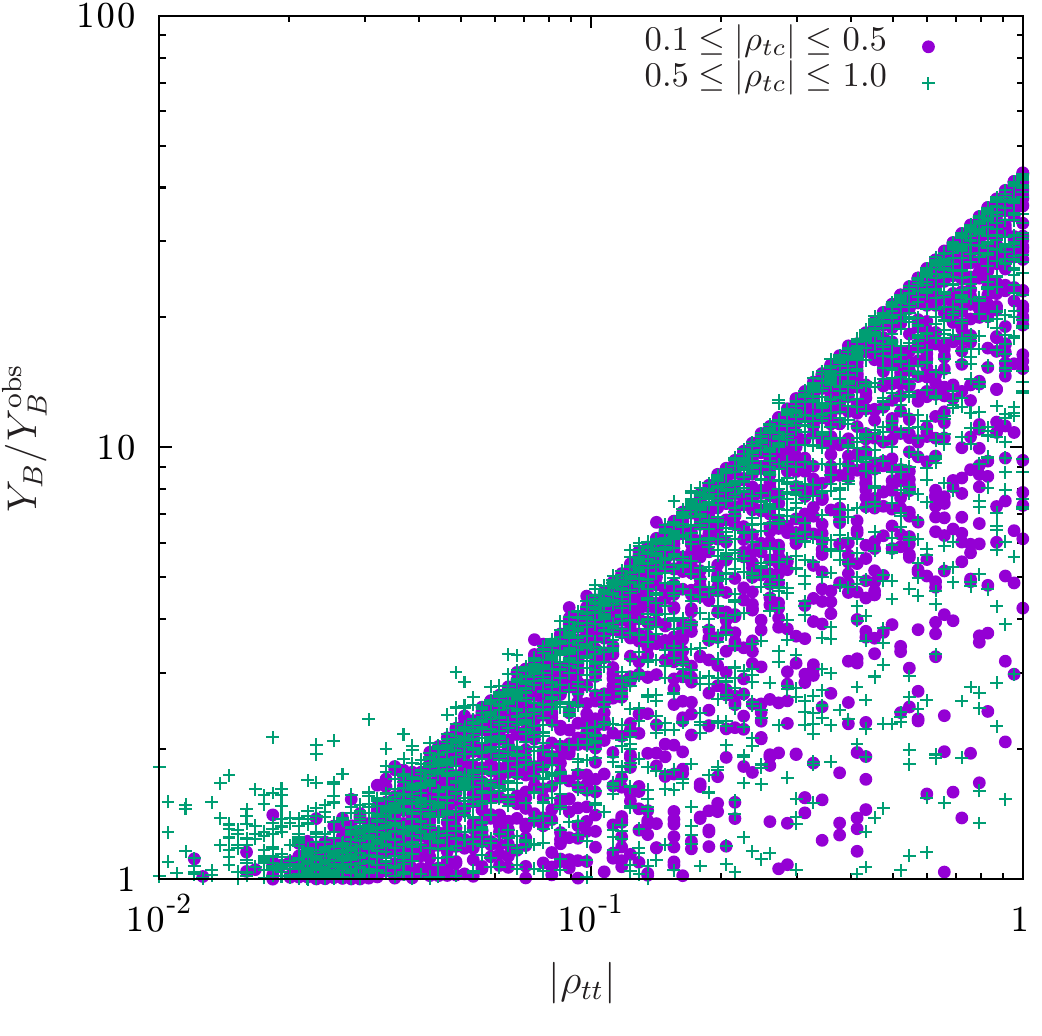}
\includegraphics[width=6.9cm]{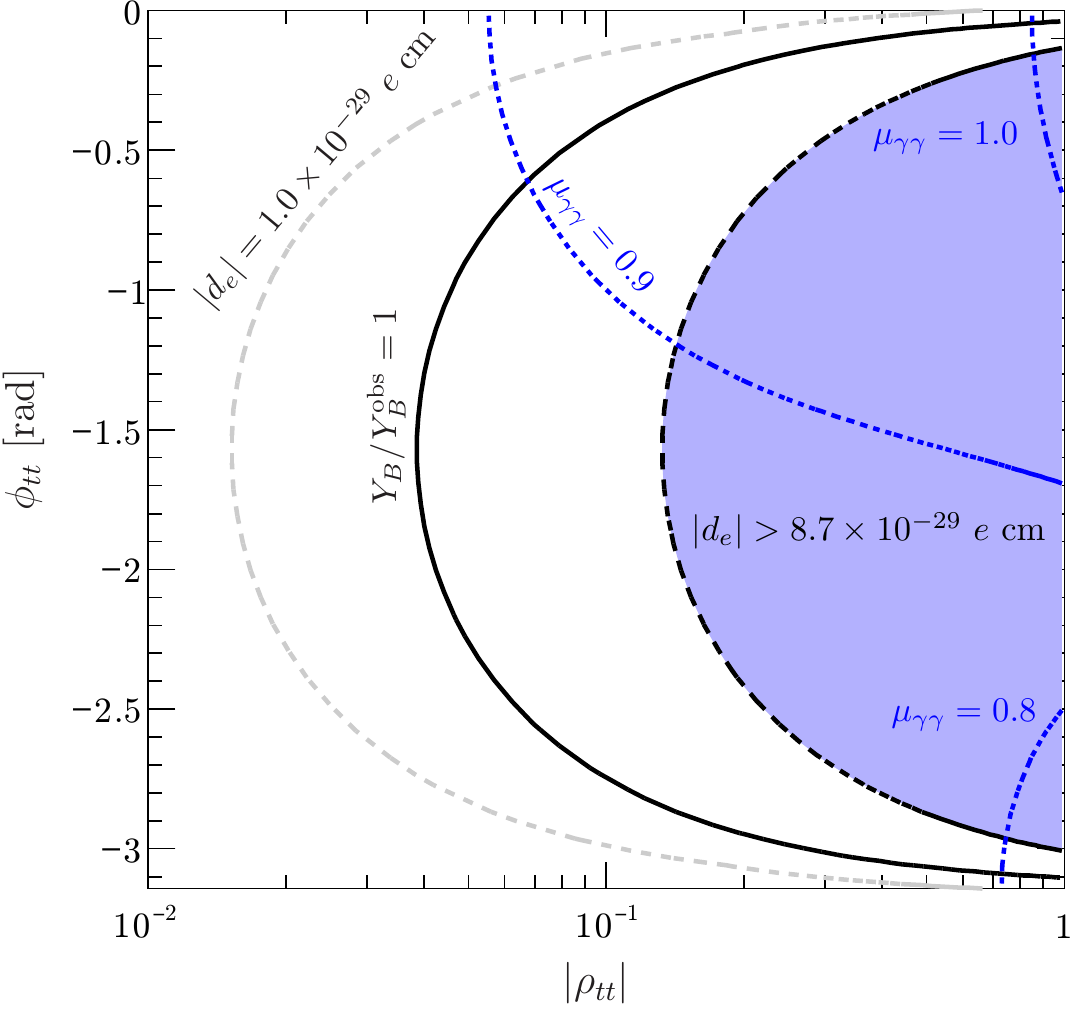}
\caption{
[left] $Y_B$ vs $|\rho_{tt}|$, where one scans over
the phases $\phi_{tt}$ and $\phi_{tc}$ $\in [0, 2\pi]$,
and purple (green) points are for 
$0.1\le |\rho_{tc}|\le 0.5$ ($0.5\le |\rho_{tc}|\le 1.0$);
[right] $Y_B$ and $|d_e|$ in 
the $|\rho_{tt}|$-$\phi_{tt}$ plane, where
dashed curve is the ACME'14 bound,
solid curve is for $Y_B = Y_B^{\rm obs}$,
and gray dashed curve is 
the ACME projected bound~\cite{ref-journal20}.
}
\label{fig:YB-EDM}
\end{figure}

To illustrate the robustness of EWBG through
Eq.~(\ref{eq:simpCPV}), we estimate $Y_B$ of
Eq.~(\ref{eq:YB}) by solving the transport equations,
where more discussion can be found in
 Refs. \cite{ref-journal13, ref-journal14}.
We plot $Y_B/Y_B^{\text{obs}}$ vs $|\rho_{tt}|$ (up to 1) 
in Fig.~\ref{fig:YB-EDM}(left), 
where we scan over $|\rho_{tc}| < 1$
and the two phases $\phi_{tt}$ and $\phi_{tc}$
(keeping physical charm and top masses), taking 
$\tan\beta = 1$ and 
$c_\gamma \equiv c_{\beta -\alpha} = 0.1$.
Furthermore, we take $m_H = m_A = m_{H^+} =$ 500~GeV
that can give rise to first order EWPT (in particular,
we have $v_ C= 176.7$ GeV $> T_C = 119.2$ 
GeV \cite{ref-journal13}),
while $\rho_{tt}$ and $\rho_{tc}$ satisfy
$B_{d,\,s}$ mixing and $b \to s\gamma$ constraints.

To discern the impact of $|\rho_{tt}|$ vs $|\rho_{tc}|$,
we separate $|\rho_{tc}|$ into lower and higher regions:
the purple dots (green crosses) are for 
$0.1 < |\rho_{tc}| < 0.5$ ($0.5 < |\rho_{tc}| < 1.0$).
For the bulk, there is no obvious distinction
between the two, so EWBG is largely $|\rho_{tt}|$-driven,
allowing $Y_B/Y_B^{\text{obs}}$ 
up to $\sim 40$ at $|\rho_{tt}| = 1$. 
However, for $|\rho_{tt}| \lesssim 0.05$,
$Y_B/Y_B^{\text{obs}}$ peters out as $|\rho_{tt}|$ drops, 
and more ``green crosses'' populate
$Y_B/Y_B^{\text{obs}} > 1$.
So, interestingly, for very small $|\rho_{tt}|$,
$\rho_{tc}$ can serve as a backup mechanism for EWBG.
But this is only possible~\cite{ref-journal13} 
for $|\rho_{tc}| = {\cal O}(1)$ with 
near maximal CPV phase, hence it is less efficient.

\subsection{Watch Your Back: {\it e}EDM} 

With $\rho_{tt}$ complex and sizable,
it can induce electron EDM through 
the two-loop mechanism \cite{ref-journal18}.
At the time of writing of Ref. \cite{ref-journal13},
May 2017, the ACME experiment had already set 
the impressively stringent limit \cite{ref-journal19} 
of $|d_e| < 8.7\times 10^{-29}\ e\,{\rm cm}$
using the polar ThO molecule,
which had to be faced. Keeping parameters as above,
where the $t \to ch$ bound at the time \cite{ref-journal4}
could be satisfied because of our low $c_\gamma$ value,
we chose to simplify by setting $\rho_{ee}$ to zero.
Setting a parameter to zero without a symmetry 
did not sound right, but we could project a range
of $d_e$ for ACME to test. 
ACME had made the projection \cite{ref-journal20} earlier
of improving by another order of magnitude,
to $1.0\times 10^{-29}\ e\,{\rm cm}$ or better.

In Fig.~\ref{fig:YB-EDM}(right) we give $Y_B$ and $|d_e|$ 
in the $|\rho_{tt}|$-$\phi_{tt}$ plane \cite{ref-journal13},
where the shaded region is ruled out by ACME'14
 \cite{ref-journal19},
the solid curve is for $Y_B/Y_B^{\text{obs}} = 1$,
and the gray dashed curve to the left
is the projected ACME bound \cite{ref-journal20}.
Little did we know that our projection had 
a shelf life of $\sim$ a year!
By October 2018, as announced in Nature \cite{ref-journal6},
\begin{equation}
  d_e < 1.1 \times 10^{-29}\; e\,{\rm cm},
      \quad {\rm (ACME\; 2018)}
 \label{eq:ACME18}
\end{equation}
the entire range of Fig.~\ref{fig:YB-EDM}(right) 
was ruled out by the ACME update.
ACME, the Advanced Cold Molecule Electron EDM experiment,
has leapt to the forefront of particle physics!

\section{Under the Heavens on Earth: {\bf e}EDM}

Having soared to the Heavens with an ${\cal O}(1)$ 
CPV source, $-\lambda_t\,\text{Im}\,\rho_{tt}$ 
as in Eq.~(\ref{eq:simpCPV}),
one has to be more attentive to $e$EDM back on Earth;
for once, Wile E. Coyote is 
keeping us honest through ACME. 
Amazingly, ACME managed to deliver on their 
projected \cite{ref-journal20} order of magnitude 
improvement within two years \cite{ref-journal6}.
It also turned out, on our side, that things were 
not as complicated or intimidating as it had seemed, 
because all loop functions for 
the two-loop mechanism are known.
We just have to turn $\rho_{ee}$ on, 
with its complex phase. 
The outcome turned out intriguing \cite{ref-journal21}.

\begin{figure}[t]
\center
\includegraphics[width=3.8cm]{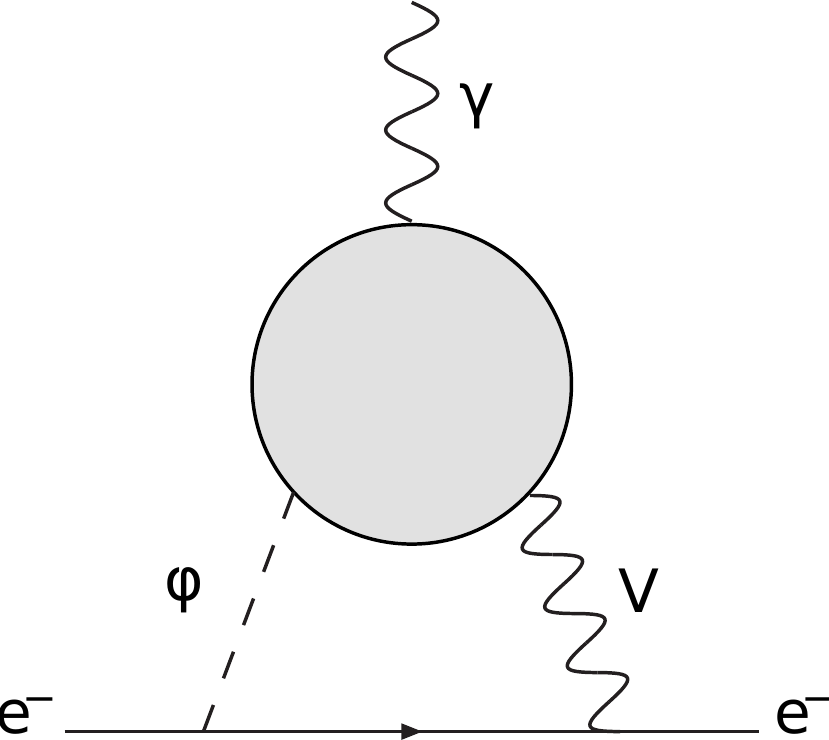}
\caption{
Two-loop Barr-Zee diagrams \cite{ref-journal18} 
contributing to the electron EDM, where 
$\phi$ denotes neutral and charged Higgs bosons,
and $V$ denotes vector bosons $\gamma$, $Z$ and $W$.
}
\label{fig:de_BZ}
\end{figure}

The dominant contributions to $d_e$ in g2HDM come from 
Barr-Zee diagrams \cite{ref-journal18}, 
as depicted in Fig.~\ref{fig:de_BZ}, which
has three pieces,
\begin{align}
d_e = d_e^{\phi\gamma} + d_e^{\phi Z} + d_e^{\phi W},
\label{eq:de}
\end{align}
where $\phi$ can be the neutral $h,H,A$ bosons
for $V = \gamma, Z$, or the $H^+$ boson for $V = W$. 
CP is violated at the lower 
and/or upper vertices of the $\phi$ line.
So, how to render $e$EDM small?
The first thing to note is that the dominant
effect comes from the ``$\phi\gamma\gamma$'' insertion
in Fig.~\ref{fig:de_BZ} for $\phi = h, H, A$, 
which is quite similar to the diagram that 
generates $h \to \gamma\gamma$,
one of the two discovery modes of $h$.

\begin{figure}[b]
\center
\includegraphics[width=3.7cm]{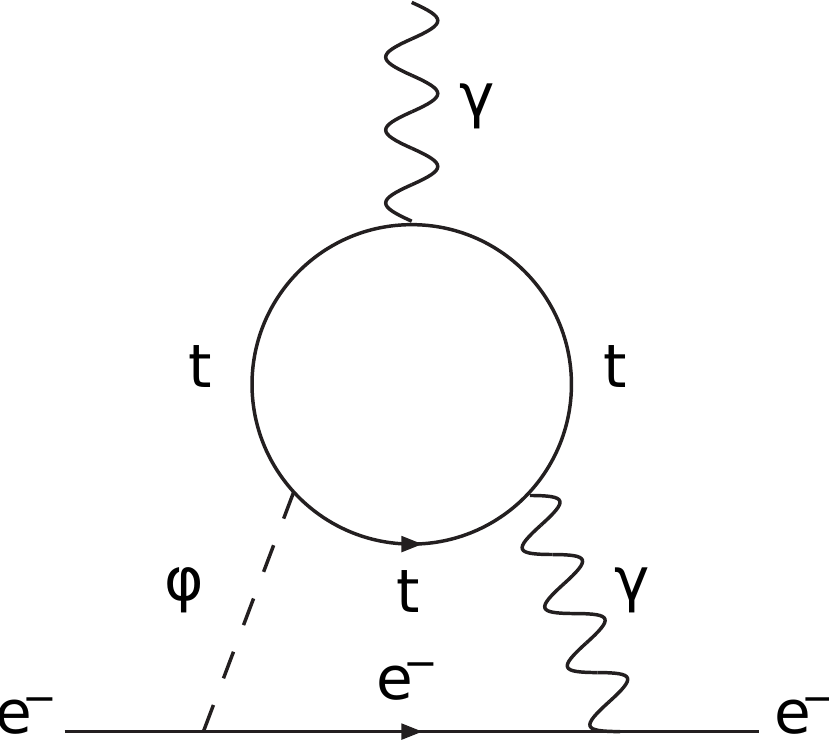}
\hspace{0.2cm}
\includegraphics[width=3.7cm]{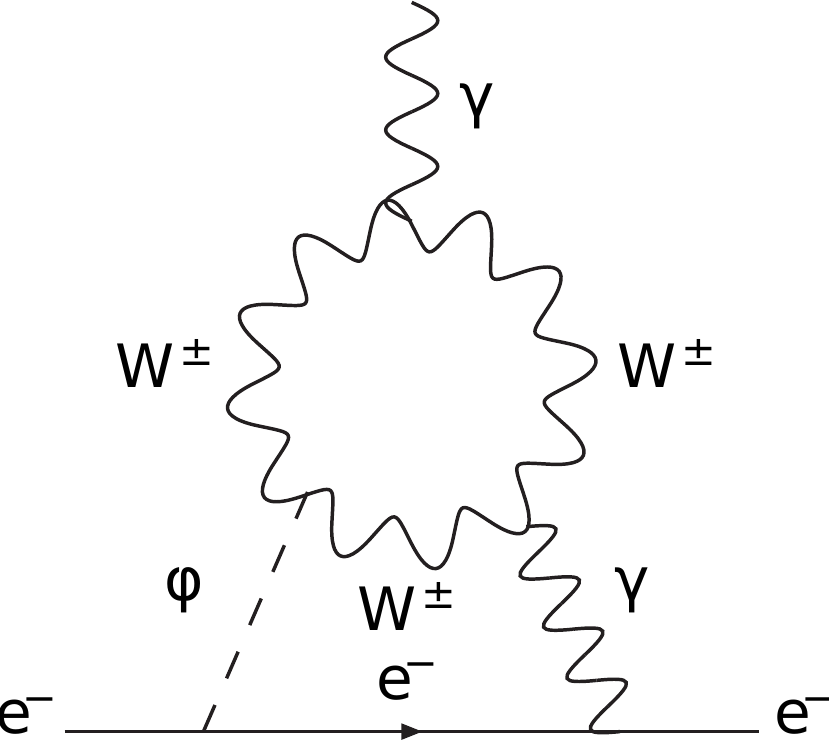}
\vspace{0.2cm}
\caption{
  Two dominant diagrams for $d_e^{\phi\gamma}$ 
  when $\text{Im}\,\rho_{ee}$ is also taken into account.}
\label{fig:de_Hgam}
\end{figure}

In Ref. \cite{ref-journal13}, we assumed 
only $\rho_{tt}$ has nonzero CPV phase, 
and set $\rho_{ee}$ in fact to zero.
Then $d_e$ is solely due to $(d_e^{\phi\gamma})_t$, which 
arises from the left diagram of Fig.~\ref{fig:de_Hgam}.
We find
\begin{align}
\frac{(d_{e}^{\phi\gamma})_{t}}{e}
= \frac{\alpha_{\rm em}s_{2\gamma}}{\,12\sqrt{2}\pi^3\,v\,}
   \frac{m_e}{m_t}\text{Im}\,\rho_{tt}\,\Delta g
= -6.6\times 10^{-29}
   \left(\frac{s_{2\gamma}}{0.2}\right)
   \left(\frac{\text{Im}\,\rho_{tt}}{-0.1}\right)
   \left(\frac{\Delta g}{0.94}\right),
\label{eq:de_Hgam_t_0}
\end{align}
where $e$ is the positron charge,
 $\alpha_{\text{em}}=e^2/4\pi$
and $\Delta g = g(m_t^2/m_h^2)-g(m_t^2/m_H^2)$, 
where the loop function $g$ is given in 
Ref.~\cite{ref-journal18}.
We have put $d_e = (d_e^{\phi\gamma})_t$
in a form to make clear that $d_e$ survives
ACME'14~\cite{ref-journal19}, but not
ACME'18~\cite{ref-journal6}.
Compared with the robust $\lambda_t\, {\rm Im}\, \rho_{tt}$ 
EWBG-driver of Eq.~(\ref{eq:simpCPV}),
the $\lambda_e\, {\rm Im}\, \rho_{tt}$ effect
of Eq.~(\ref{eq:de_Hgam_t_0}) did not pass ACME scrutiny.

%
So, to survive ACME'18, one needs 
a cancellation mechanism for $d_e^{\phi\gamma}$,
upon turning on $\rho_{ee}$ with its CPV phase.

\subsection{Cancellation Mechanism for Electron EDM}

For a cancellation mechanism, one naturally
recalls the cancellation between 
the top and $W$ loops for $h \to \gamma\gamma$,
the diphoton decay of $h$, 
where in fact the $W$ loop dominates over top.
Upon turning on $\text{Im}\,\rho_{ee}$, 
a similar effect can happen to make 
$(d_e^{\phi\gamma})_W$ of Fig.~\ref{fig:de_Hgam}(right)
comparable to or even bigger than
$(d_e^{\phi\gamma})_t$ of Fig.~\ref{fig:de_Hgam}(left).

Let us separate $(d_{e}^{\phi\gamma})_{i}$ into two parts
\begin{align}
  (d_{e}^{\phi\gamma})_i \equiv (d_{e}^{\phi\gamma})_{i}^{\text{mix}}
 +(d_{e}^{\phi\gamma})_{i}^{\text{extr}}.
\end{align}
The first term comes from mixing both 
SM and extra Yukawa couplings, while 
the second term involves extra Yukawa couplings only.
From Fig.~\ref{fig:de_Hgam}(left), 
one has for the top-loop
\begin{align}
\frac{(d_{e}^{\phi\gamma})_{t}^{\text{mix}}}{e}
&= \frac{\alpha_{\rm em}s_{2\gamma}}{12\sqrt{2}\pi^3\,v\,}
   \left[\text{Im}\,\rho_{ee}\,\Delta f
	 +\frac{m_e}{m_t}\text{Im}\,\rho_{tt}\,\Delta g\right],
 \label{eq:de_Hgam_t} \\
   \frac{(d_{e}^{\phi\gamma})_{t}^{\text{extr}}}{e}
&\simeq \frac{\alpha_{\text{em}}}{12\pi^3\,m_t\,}
 \text{Im}(\rho_{ee}\,\rho_{tt})
   \Big[f(\tau_{tA})+g(\tau_{tA})\Big],
 \label{eq:de_Hgam_t_ex}
\end{align}
where {$\tau_{ij}=m_i^2/m_j^2$, $\Delta X = X(\tau_{th})-X(\tau_{tH})$,}
and $X=f,\,g$ are monotonically increasing 
loop functions given in Ref.~\cite{ref-journal18};
thus $\Delta X >0$ for $m_h < m_H$.
Note that Eq.~(\ref{eq:de_Hgam_t}) is 
an extension of Eq.~(\ref{eq:de_Hgam_t_0})
for $\text{Im}\,\rho_{ee} \neq 0$,
while Eq.~(\ref{eq:de_Hgam_t_ex}) depends on
the phase difference between $\rho_{ee}$ and $\rho_{tt}$.
We have made the approximation of
$c_\gamma^2 \ll 1$ and $m_H\simeq m_A$ to simplify 
the appearance of $(d_{e}^{H\gamma})_{t}^{\text{extr}}$, 
but this is not imposed in our later numerics.

For the $W$-loop of Fig.~\ref{fig:de_Hgam}(right), 
the $\phi WW$ vertex involves SM couplings 
modulated by $h$-$H$ mixing,
so $(d_e^{\phi\gamma})_W$ is solely given by $(d_{e}^{\phi\gamma})_W^{\text{mix}}$, 
\begin{align}
 \frac{(d_{e}^{\phi\gamma})_W^{\text{mix}}}{e}
& = -\frac{\alpha_{\text{em}}s_{2\gamma}} 
          {64\sqrt{2}\pi^3\,v\,}
     \text{Im}\,\rho_{ee}\,\Delta\mathcal{J}^\gamma_W,
 \label{eq:de_Hgam_W}
\end{align}
where $\Delta\mathcal{J}^\gamma_W
 = \mathcal{J}^\gamma_W(m_h)-\mathcal{J}^\gamma_W(m_H)$.
The function $\mathcal{J}^\gamma_W$ is given in 
Ref.~\cite{ref-journal22}, which is monotonically decreasing,
hence $\Delta\mathcal{J}^\gamma_W > 0$ for $m_h < m_H$.

To suppress $d_e^{\phi\gamma}$, we consider
the cancellation between top and $W$ loops
for the $h$-$H$ mixing terms,
$(d_e^{\phi\gamma})_t^{\text{mix}} +  (d_e^{\phi\gamma})_W^{\text{mix}} = 0$,
while the purely extra Yukawa term vanishes,
$(d_e^{\phi\gamma})_t^{\text{extr}} = 0$.
We will discuss 
$(d_e^{\phi\gamma})_t^{\text{extr}} \neq 0$ later.
Comparing Eqs.~(\ref{eq:de_Hgam_t}) and (\ref{eq:de_Hgam_W})
for the first condition, and Eq.~(\ref{eq:de_Hgam_t_ex})
for the second, one finds
\begin{align}
 \frac{\text{Im}\,\rho_{ee}}{\text{Im}\,\rho_{tt}}
 = c\, \frac{\lambda_e}{\lambda_t},\quad\
 \frac{\text{Re}\,\rho_{ee}}{\text{Re}\,\rho_{tt}}
 = -\frac{\text{Im}\,\rho_{ee}}{\text{Im}\,\rho_{tt}},
\label{eq:de_cancel_phigam}
\end{align}
respectively, where $c = (16/3)\Delta g/
[\Delta\mathcal{J}_W^\gamma-(16/3)\Delta f]$.
For example, $c\simeq 0.71$ 
for $m_h, m_H = 125$, 500~GeV.
Combining the two conditions of
Eq.~(\ref{eq:de_cancel_phigam}), one gets
\begin{equation}
 \left|\frac{\rho_{ee}}{\rho_{tt}}\right|
  = c \, \frac{\lambda_e}{\lambda_t},
  \quad\
 \text{Im}(\rho_{ee}\,\rho_{tt}) \to 0,
\label{eq:correl}
\end{equation}
with correlated phase between 
$\rho_{tt}$ and $\rho_{ee}$ as indicated.
Note that $c$ is not sensitive to the detailed
exotic Higgs spectrum that is consistent with 
first order EWPT, hence does not change drastically 
in the parameter range for EWBG.

\subsection{Facing ACME: Thorium Oxide EDM}

Having elucidated the cancellation mechanism
for the dominant $d_e^{\phi\gamma}$ term,
to understand our more detailed numerics,
we need to include the subdominant effects of
$d_e^{\phi W}$ and $d_e^{\phi Z}$ in Eq.~(\ref{eq:de}).
We also need to tune in to a little more
detail in making contact with experiment, 
i.e. how the measurement is actually done.

As ThO is a polar molecule with very strong
internal electric field, we need to understand
some ``environment'' effects. 
The effective EDM for ThO is \cite{ref-journal23}
\begin{align}
 d_{\text{ThO}} = d_e+\alpha_{\text{ThO}}C_S,
\label{eq:dThO}
\end{align}
where $d_e$ is the coefficient of the dimension-5 operator
$-\frac{i}{2}d_e(\bar{e}\sigma^{\mu\nu}\gamma_5e)F_{\mu\nu}$,
 with $F_{\mu\nu}$ the EM field strength tensor.
The $C_S$ term is due to 
T-violating electron-nucleon interaction, 
$-\frac{G_F}{\sqrt{2}}C_S(\bar{N}N)(\bar{e}i\gamma_5 e)$,
with $G_F$ the Fermi constant.
The ACME'18 result of Eq.~(\ref{eq:ACME18})
corresponds to
 $d_{\text{ThO}} = (4.3 \pm 4.0) \times 10^{-30}\; e$\, cm,
but taking $C_S = 0$ \cite{ref-journal6}.
For our case, an estimate \cite{ref-journal24} of $\alpha_{\text{ThO}}=1.5\times10^{-20}$
implies $C_S$ cannot be neglected w.r.t.
$d_e^{\phi Z}$ and $d_e^{\phi W}$ of Eq.~(\ref{eq:de}), 
and we use $d_{\text{ThO}}$ of ACME'18 
to explore the constraint.

We follow the estimate for $C_S$ from Ref.~\cite{ref-journal25}
(consistent with Ref.~\cite{ref-journal26}),
\begin{align}
C_S = -2v^2 \bigg[6.3\,(C_{de}+C_{ue})
    + C_{se}\frac{41~\text{MeV}}{m_s}
	+ C_{ce}\frac{79~\text{MeV}}{m_c} 
	+ 0.062 \left(\frac{C_{be}}{m_b}
	            + \frac{C_{te}}{m_t}\right)\bigg],
 \label{eq:CS}
\end{align}
where $C_{qe}$ is given by 
$\mathcal{L}_{4f}^{\text{CPV}} = 
 \sum_qC_{qe}(\bar{q}q)(\bar{e}i\gamma_5 e)$
after integrating out all neutral Higgs bosons.
Note that quark masses are balanced
  by corresponding Yukawa couplings in $C_{qe}$,
  so all quark flavors are generally relevant.
Note also that for $s_\gamma \to 1$ and $m_H\simeq m_A$, 
one has
$C_{ue}\simeq \text{Im}(\rho_{ee}\rho_{uu})/(2m_A^2)$ and
$C_{de} \simeq \text{Im}(\rho_{ee}\rho_{dd}^*)/(2m_A^2)$, respectively, implying that $C_{qe}\simeq 0$ for $(d_e^{\phi\gamma})_{q}^{\text{extr}}\simeq 0$.

\begin{figure*}[t]
\center
\includegraphics[width=6.85cm]{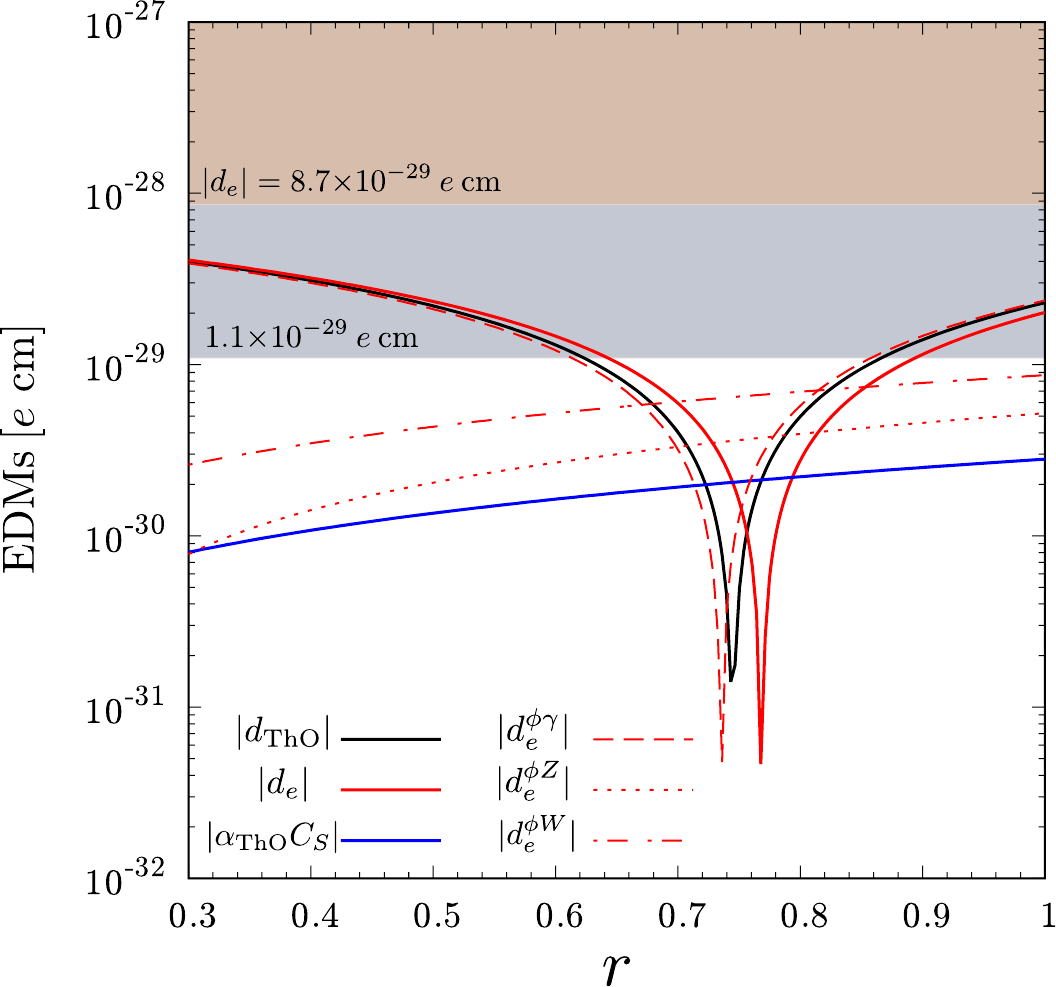}
\hspace{0.25cm}
\includegraphics[width=6.4cm]{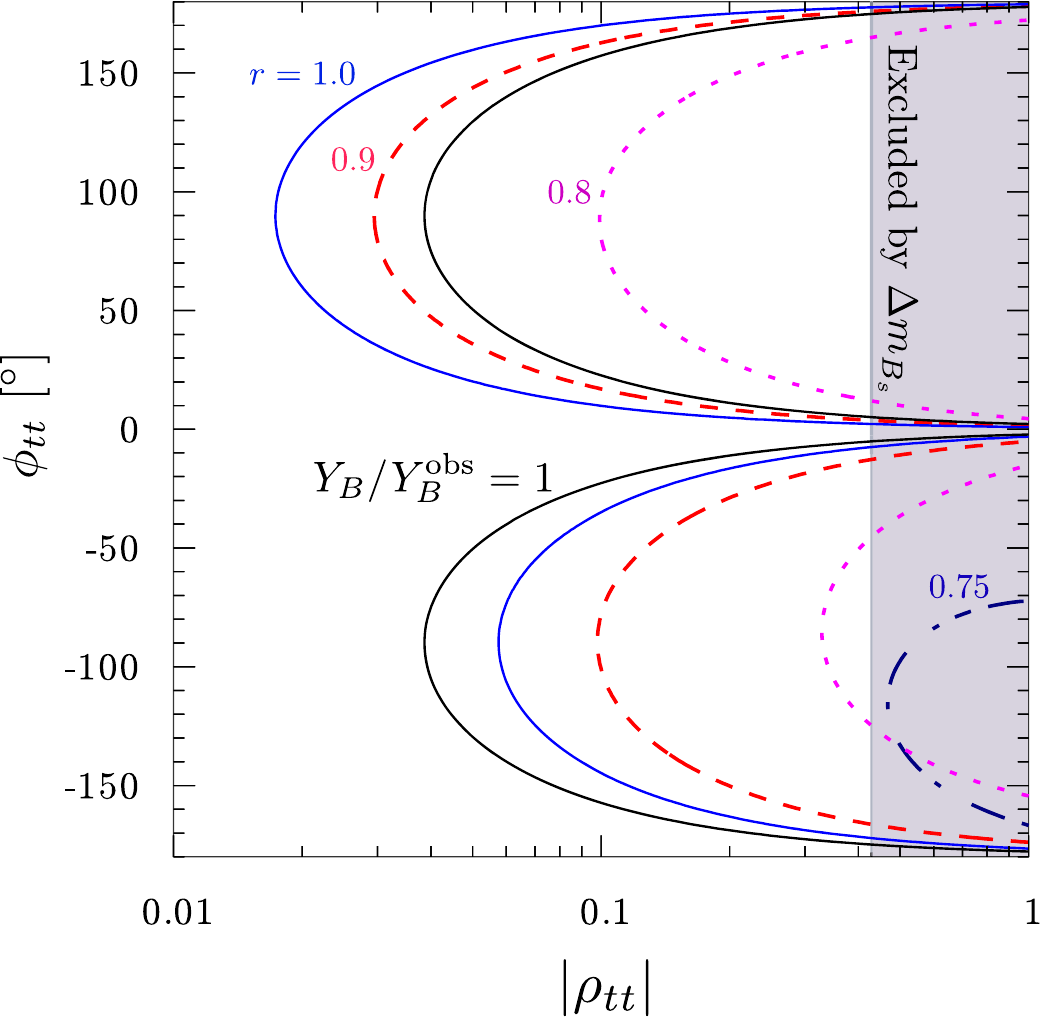}
\caption{
[Left] $|d_{\rm ThO}|$ and its components 
 vs $r$, as defined in Eq.~(\ref{eq:Ansatz}). 
 We take $\text{Re}\,\rho_{tt}=\text{Im}\,\rho_{tt}=-0.1$,
 $c_\gamma=0.1$ and $m_H=m_A=m_{H^\pm}=500$ GeV,
 with ACME bounds overlaid.
[Right] 2$\sigma$-allowed region of $d_{\text{ThO}}$ 
 with $r=1.0$ (blue, solid), 0.9 (red, dashed), 
 0.8 (magenta, dotted) and 0.75 (navy blue, dot-dashed),
 respectively.
 The region to the right of the black solid contour, $Y_B/Y_B^{\text{obs}}=1$, is allowed. 
 The gray shaded region is nominally excluded 
 by $B_s$-$\bar{B}_s$ mixing. 
}
\label{fig:dThO}
\end{figure*}

We find~\cite{ref-journal21} one-loop induced 
CPV mixing of neutral Higgs bosons to be minor, 
so we turn to the numerics.
Motivated by Eq.~(\ref{eq:de_cancel_phigam}),
we consider the simplified ``Ansatz'',
\begin{align}
\frac{\text{Im}\,\rho_{ff}}{\text{Im}\,\rho_{tt}}
  = \ \,\, r\, \frac{\lambda_f}{\lambda_t}, \quad 
\frac{\text{Re}\,\rho_{ff}}{\text{Re}\,\rho_{tt}}
  = -r\,  \frac{\lambda_f}{\lambda_t},
\label{eq:Ansatz}
\end{align}
which is flavor-blind.
Keeping EWBG in mind, we choose $\rho_{tt}$
strength that is able to drive it.
In Fig.~\ref{fig:dThO}(left) we plot
 $|d_{\text{ThO}}|$ (black, solid) and its components
 $|d_e|$ (red, solid),
 $|d_e^{\phi\gamma}|$ (red, dashed),
 $|d_e^{\phi W}|$ (red, dot-dashed), 
 $|d_e^{\phi Z}|$ (red, dotted), 
 $|\alpha_{\text{ThO}}C_S|$ (blue, solid) 
as functions of $r$, where we set $\text{Re}\,\rho_{tt}=\text{Im}\,\rho_{tt}=-0.1$, 
$c_\gamma=0.1$, and common $H$, $A$, $H^+$ mass
at 500~GeV for illustration, which we will 
show below that it can generate successful EWBG.
The ACME'18~\cite{ref-journal6} 
and previous (ACME'14) bounds are  
the {gray and light brown} shaded regions. 
The absence of $\rho_{ee}$ corresponds to $r=0$, 
with $d_e\simeq (d_e^{\phi\gamma})_t$ 
estimated in Eq.~(\ref{eq:de_Hgam_t_0}).
This specific point \cite{ref-journal13}, 
far to the left and outside the plot, 
is excluded by ACME'18. 
However, the situation is considerably 
different~\cite{ref-journal21} for $r\neq0$.

We see strong cancellation in $d_e^{\phi\gamma}$
around $r \simeq 0.75$, as mentioned in 
our discussion after Eq.~(\ref{eq:de_cancel_phigam}).
This is due to $(d_e^{\phi\gamma})_W$ of 
Fig.~\ref{fig:de_Hgam}(right) canceling against
Fig.~\ref{fig:de_Hgam}(left) for $\text{Im}\,\rho_{ee} \neq 0$.
The subdominant effects of $d_e^{\phi W}$ and $d_e^{\phi Z}$ 
then shift the cancellation point for $d_e$ upwards in $r$.
Finally, to get $d_{\rm ThO}$, the $C_S$ contribution 
needs to be added, which moves $d_{\rm ThO}$ 
downwards a bit. The upshot is that, 
owing to this cancellation mechanism, 
$d_{\text{ThO}}$ can be suppressed by two orders 
of magnitude below the ACME'18 bound.
We remark that, in principle $r$ can depend on
the fermion flavor $f$ (see Eq.~(\ref{eq:Ansatz})),
enriching the possibilities. On the other hand, 
even finding a preferred $r$ value in the future,
it depends on several loop functions with
various input parameters to disentangle.

With electron EDM under control and with 
a lot of leeway to face experimental scrutiny,
it is imperative now to check whether EWBG survives.
The 2$\sigma$ allowed region of $d_{\text{ThO}}$ 
is displayed in Fig.~\ref{fig:dThO}(right) in the $|\rho_{tt}|$--$\phi_{tt}$ plane, for
$r=1.0$ (blue, solid), 0.9 (red, dashed), 
0.8 (magenta, dotted) and 0.75 (navy blue, dot-dashed).
Regions to the left of these contours are allowed, 
while to the right of the black contours correspond to
$Y_B > Y_B^{\text{obs}} = 8.59\times 10^{-11}$,
consistent with Planck 2014~\cite{ref-journal15} for EWBG.
The gray shaded region for larger $|\rho_{tt}|$ values
is excluded by $B_s$-$\bar{B}_s$ mixing, but 
this is by ignoring tree diagrams due to $\rho_{sb}$, etc.
In Ref.~\cite{ref-journal13} we considered 
$\phi_{tt} < 0$ for BAU positive.
However, one can flip the sign of $\Delta\beta$ 
to get $\phi_{tt} > 0$. 
Since the central value of $d_{\text{ThO}}$ is positive,
the allowed region is asymmetric in $\phi_{tt}$.
For $r=1.0$ and 0.9, only $\phi_{tt} < 0$ is 
consistent with $\rho_{tt}$-driven EWBG.
But $\phi_{tt} > 0$ becomes possible as $r$ approaches 
the cancellation point at $r \sim 0.75$, 
enlarging the solution space for $\rho_{tt}$-driven EWBG.

Our results presented here are illustrative and 
do not exhaust the parameter space that
satisfy both EWBG and $e$EDM.
We remark that a cancellation mechanism would 
not be necessary if $\rho_{tt}$ is very small 
but EWBG is driven by $\rho_{tc}$.

\subsection{The Flavor Enigma and NFC}


Having survived the $e$EDM bound of ACME'18,
and with EWBG once again demonstrated,
we cannot help but exclaim 
(quoting a Psalm of David):

\begin{center}
\hskip-1.46cm O Lord, our Lord, \\
   \hskip0.75cm How Majestic is Thy Name \\
   \hskip0.8cm  in all the Earth, \\
   \hskip0.8cm Who have set Thy Splendor \\
   \hskip0.9cm  above the Heavens !
\end{center}
There may be {\it something} to g2HDM.

Our simplified Ansatz to achieve cancellation, Eq.~(\ref{eq:Ansatz}),
basically says that the diagonal elements of 
the extra Yukawa couplings $\rho_{ff}$ 
follow the same SM hierarchy,
with specific correlation of CPV phases
between charged leptons and up-type quarks.
Although the Ansatz enters also $C_S$, the most 
relevant couplings are $\rho_{tt}$ and $\rho_{ee}$,
reflecting the largest and smallest Yukawa couplings
$\lambda_t$ and $\lambda_e$, which differ 
by six orders of magnitude!

We now turn to comment on how the NFC condition 
of Glashow and Weinberg is a prejudice in itself.
At the time of their paper, they new very well
the fermion mass hierarchy, namely
\begin{align}
 m_e \ll &\ m_\mu \ll m_\tau, \nonumber \\
 m_d \ll &\ m_s \ll m_b, \nonumber \\  
 m_u \ll &\ m_c \ll m_t, 
\label{eq:ma-hier}
\end{align}
between the fermion generations.
What they did not know was how heavy 
the top quark actually turned out to be.
That is, nobody anticipated the very large ratio
\begin{equation}
 m_t/m_b \gg 1,
\label{eq:tob}
\end{equation}
as attested by the parade of accelerators,
PEP, PETRA, TRISTAN, each holding hope
to be the one to capture the top quark.
But it was the ARGUS discovery~\cite{ref-journal27} 
of large $B^0$-$\bar B^0$ mixing in 1987 that 
harbingered the heaviness of the top and spelled out 
the null search prospect at SLC/LEP beforehand.
It took almost another decade for the top to be
discovered~\cite{ref-journal4} at the Tevatron.
The large $B^0$-$\bar B^0$ mixing, however,
sowed the seed for the eventual B factories
that confirmed the Kobayashi-Maskawa CPV mechanism 
of the SM quark sector.

If the fermion mass hierarchy of Eq.~(\ref{eq:ma-hier})
was known in 1977, except the very large $m_t/m_b$
ratio (much larger than $m_c/m_s$), what came 
totally out of whack, circa 1983~\cite{ref-journal4}
is the quark mixing hierarchy,
\begin{equation}
    V_{ub}^2 \ll V_{cb}^2 \ll V_{us}^2 \ll V_{tb}^2 \cong 1,
\label{eq:mi-hier}
\end{equation}
which was totally an experimental discovery
(MAC/Mark II for long $b$ lifetime, and
 CLEO for absence of $b \to u$) 
and not anticipated at all.
Note that $V_{us}^2 \sim 1/20 \ll 1$ 
was known for a long time, 
which led to the naming of ``strangeness''.
But the quark mixing hierarchy 
of Eq.~(\ref{eq:mi-hier}) came out of the blue, 
which led Wolfenstein to propose his namesake 
parametrization~\cite{ref-journal28},which influenced 
the placing~\cite{ref-journal4} of the KM phase.

With the mass-mixing hierarchies 
of Eqs.~(\ref{eq:ma-hier}) and (\ref{eq:mi-hier}),
probably stimulated by the ARGUS suggestion 
for the {\it heaviness} of the top, Eq.~(\ref{eq:tob}), 
Cheng and Sher suggested~\cite{ref-journal29}, 
a decade after the Glashow-Weinberg paper, 
that one may not need NFC if FCNH couplings 
trickle down off-diagonal $\propto \sqrt{m_i m_j}/v$.
A few years later, we pointed out~\cite{ref-journal30} 
that, based on the Cheng-Sher Ansatz,
the decay mode to watch is $t \to ch$,
if some Higgs boson $h$ is lighter than top,
or $h \to t\bar c$ if $h$ is heavier.
The paper also stressed that the Cheng-Sher Ansatz
was too specific, and one can more broadly define
a ``2HDM III'' just based on
the quark mass-mixing hierarchies.
This 2HDM III we now call g2HDM, 
or even SM2 --- SM with two scalar doublets,
as the mass-mixing hierarchies are 
``known'' to both doublets.

Nature's design of fermion mass-mixing hierarchies
--- the flavor enigma --- seems to hide well
the extra Higgs bosons $H$, $A$, $H^+$ to this day.
That the extra Yukawa couplings would echo
this observed mass-mixing hierarchy structure,
as exemplified in Eq.~(\ref{eq:Ansatz})
that evades ACME probe with $e$EDM, 
with a couple of order of magnitudes to spare,
seems staggering. 
Even more perplexing is that Nature
seemingly sent in another unexpected ``protection'': 
the emergent phenomenon of small $c_\gamma$, 
alignment, that can handily explain 
the absence of $t \to ch(125)$ so far, 
without requiring $\rho_{tc}$ to be much less than 1.

At what length would Nature go to hide from us these 
extra Higgs bosons and their associated couplings?

\subsection{Comments: On ``the Heavens and the Earth''}

\vskip0.12cm\hskip1.75cm\parbox{3.28in}
 {In context of EWBG driven by an {\it extra top 
 Yukawa} coupling, the impressive ACME'18 bound 
 suggests an {\it extra electron Yukawa} coupling 
 that works in concert to give {\it exquisite 
 cancellation} among dangerous diagrams.  
 The cancellation mechanism calls for the 
 {\it extra Yukawas} to {\it echo} the 
 hierarchical {\it pattern of SM Yukawa}
 couplings.}\vskip0.25cm
 
As we quoted from Psalm 8, of David, which echoes 
our Title, the theme of this article is to strike 
the contrast between
--- {\bf the Heavens}, and {\bf the Earth}.
To explain the disappearance of antimatter from
the early Universe, or the Heavens,
one needs lofty, new CPV phases that are
very BSM, i.e. beyond the KM phase.
But with such large CPV phase,
can one survive the very stringent scrutiny
from LE precision frontier probes, such as ACME?
The g2HDM provides an existence proof, which
also illustrates the delicateness.
The contrast between $\rho_{tt}$ and $\rho_{ee}$,
that it echoes the $\lambda_t$ vs $\lambda_e$
pattern, is truly curious.

A second point to make is that, 
ACME'14 was basically confirmed by 
the JILA group~\cite{ref-journal31} 
using trapped HfF$^+$ ions.
This is a different approach from ACME, hence 
fulfills the standard criterion of an independent check.
However, the ACME'18 result has not been confirmed
independently so far, although several groups
are galvanized to join the fray. 
In this sense, we do not view $10^{-29}\;e$\,cm
as ``finished business'', for we have
witnessed, in our own lifetime, 
{\it discovery}~\cite{ref-journal27} right on top of
the previously set~\cite{ref-journal32} bound!

It is really amusing that the {\it largest} 
diagonal extra Yukawa $\rho_{tt}$ drives B.A.U.,
while working in concert with the {\it smallest}
diagonal extra Yukawa $\rho_{ee}$ to generate $e$EDM,
which might be revealed soon by 
very-low-energy ultra-precision probes. 
It should be clear that the
$10^{-29}$ to $10^{-30}\;e$\,cm range
seems ripe and fabulous.
Godspeed their success!

What lies between $\rho_{tt}$ and $\rho_{ee}$,
spanning 6 orders of magnitude, the various
extra Yukawa couplings provide a host of 
phenomena that can be probed at the LHC 
as well as the {\it flavor} frontier,
to which we now turn.

\section{Phenomenological Consequences}

As the extra Yukawa couplings reflect 
the SM Yukawa pattern, we shall 
take~\cite{ref-journal33,ref-journal34}
the conservative
\begin{equation}
 \rho_{ii} \lesssim {\cal O}(\lambda_i); \ \ \ 
 \rho_{1i} \lesssim {\cal O}(\lambda_1); \ \ \ 
 \rho_{3j} \lesssim {\cal O}(\lambda_3)  \ (j \neq 1), 
\label{eq:pattern}
\end{equation}
since the bounds involving the third generation 
are the weakest.
The main phenomenological consequences are
searching for $H$, $A$, $H^+$ at the LHC,
and combing the flavor frontier; 
our discussion will be brief.
We shall also comment on the need of
${\cal O}(1)$ Higgs quartics, namely 
the issue of order of phase transition,
and implications on the Landau ghost.
On the side, we comment on the possible impact of 
the recently confirmed muon $g-2$ anomaly.

\begin{figure}[h]
\center
\includegraphics[width=.47\textwidth]{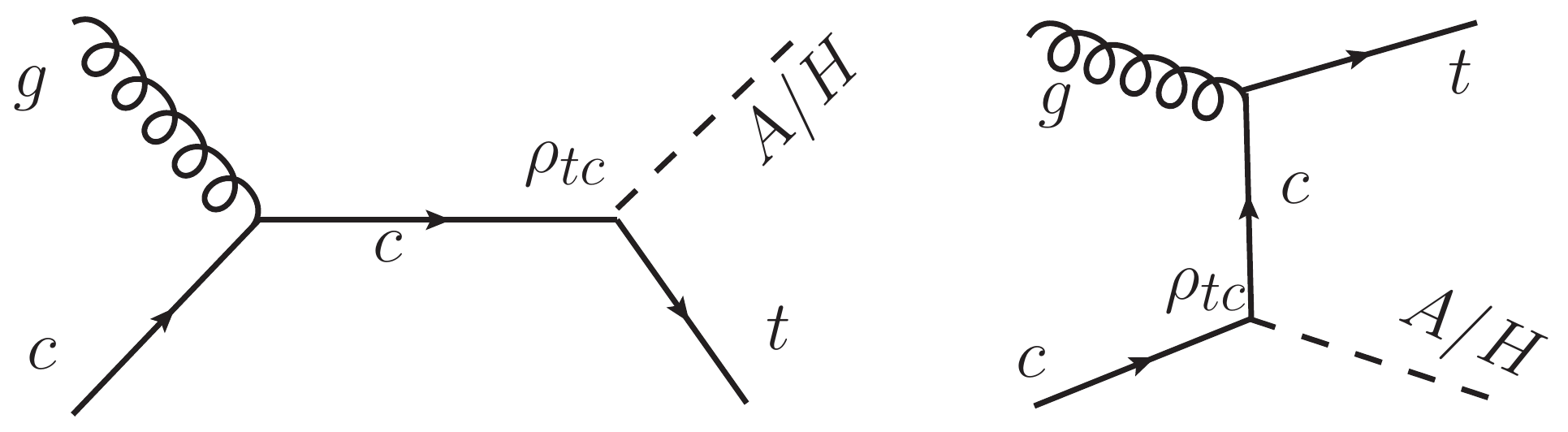}
 \hskip0.5cm
\includegraphics[width=0.4\textwidth]{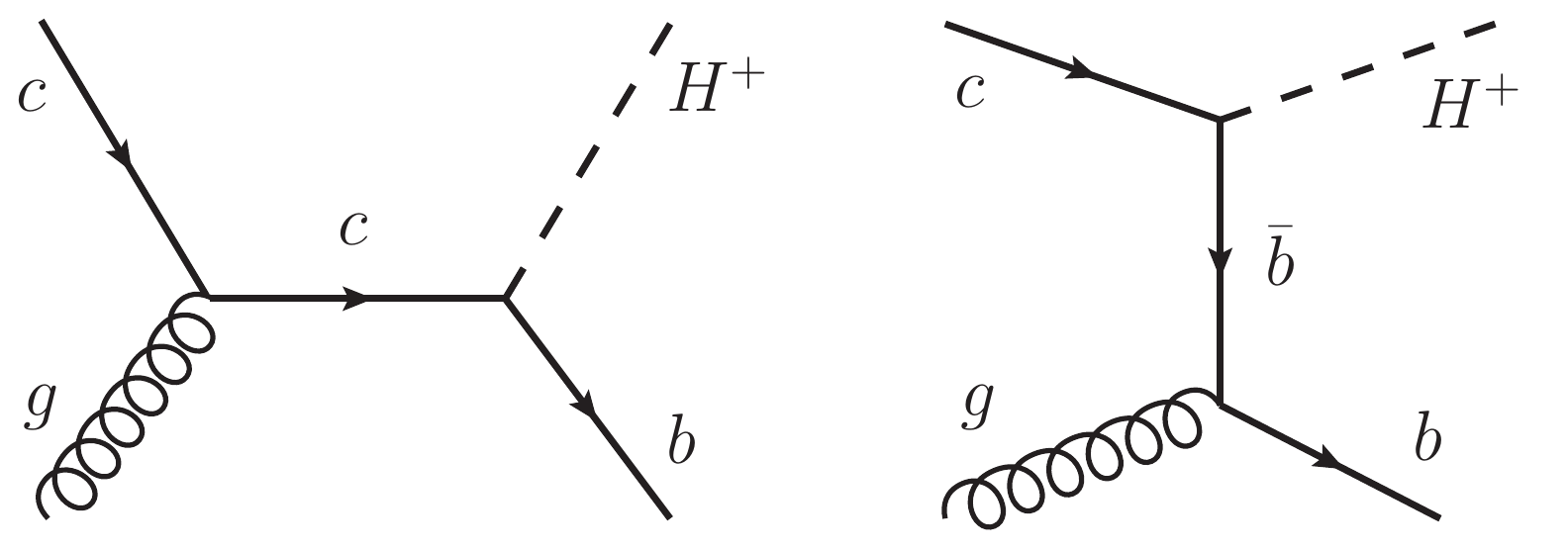}
\caption{
 Feynman diagrams for the $cg \to tA,\,tH$
 and $cg\to b H^+$ associated production processes.
}
\label{fig:tAH-bH+}
\end{figure}

\subsection{Leading Search Modes at the LHC}

Shortly after our EWBG~\cite{ref-journal13} 
and alignment~\cite{ref-journal12} studies,
we capitalized on the $\rho_{tc}$ and $\rho_{tt}$
couplings and proposed~\cite{ref-journal35} 
the associated production mechanism, i.e. the two 
diagrams on the left side of Fig.~\ref{fig:tAH-bH+},
\begin{align}
 cg \to tH/tA \to 
 & \ tt\bar c, \label{eq:tHA} \\
 & \ tt\bar t, \nonumber
\end{align}
where production is due to $\rho_{tc}$,
while $A/H$ decay can go through 
$\rho_{tc}$ or $\rho_{tt}$, giving rise 
to $tt\bar c$ (Same-Sign Top plus $c$-jet),
or $tt\bar t$ (Triple-Top) signatures.
The discovery potential for $tt\bar c$ already looks 
promising~\cite{ref-journal35} with LHC Run 2 data, 
while the more exquisite Triple-Top,
at higher threshold and with tiny SM cross section, 
can be explored at the High-Luminosity LHC (HL-LHC).
See Ref.~\cite{ref-journal36} for further discussion.
Note that $4t$ has a SM cross section 
at ${\cal O}(12)$\,fb, which is about 
an order of magnitude larger than triple-top in SM,
and has been fervently searched for by both ATLAS 
and CMS, which puts constraints~\cite{ref-journal37} 
on $\rho_{tc}$ and $\rho_{tt}$.

The down-type Yukawa interaction is analogous
to Eq.~(\ref{eq:Yuk_u}), while the corresponding
charged Higgs Yukawa coupling can be found
in Ref.~\cite{ref-journal36}.
It is curious that it took some while for us
to come up with~\cite{ref-journal38} 
the novel $H^+$ associated production
process illustrated in the two diagrams 
on the right side of Fig.~\ref{fig:tAH-bH+},
i.e.
\begin{equation}
 cg \to bH^+ \to bt\bar b,
\label{eq:bH+}
\end{equation}
where $\rho_{tc}$ enters the $\bar cbH^+$ coupling,
while $H^+ \to t \bar b$ goes through $\rho_{tt}$.
It is not surprising that the latter coupling has
an associated $V_{tb}$ CKM factor, but a bit 
counter-intuitively (compared with 2HDM~II), 
the former also has the $V_{tb}$ CKM factor 
rather than $V_{cb}$, hence it is enhanced by 
$V_{tb}/V_{cb}$ at amplitude level~\cite{ref-journal38} . 
Furthermore, association with the $b$ quark means the 
threshold is lower than $tH,\,tA$ production,
hence it is generally more efficient.

Our study~\cite{ref-journal38} did not find severe 
backgrounds for the $bt\bar b$ signature,
but we await experimental scrutiny to find out
whether there are some yet unspecified background.
With $\rho_{tc}$ and $\rho_{tt}$ largely
unexplored, and keeping in mind each one
of them could possibly drive EWBG, we
urge the ATLAS and CMS experiments to
make a serious effort to search for the
processes of Eqs.~(\ref{eq:tHA}) and (\ref{eq:bH+}).
We note that, even if the signatures are discovered, 
reconstructing the $H$, $A$ and $H^+$ bosons would 
be the next challenge, let alone disentangling 
the CPV phase of $\rho_{tt}$ down the line.

\begin{figure*}[h]
\centering
\includegraphics[angle=0,width=13.7cm]{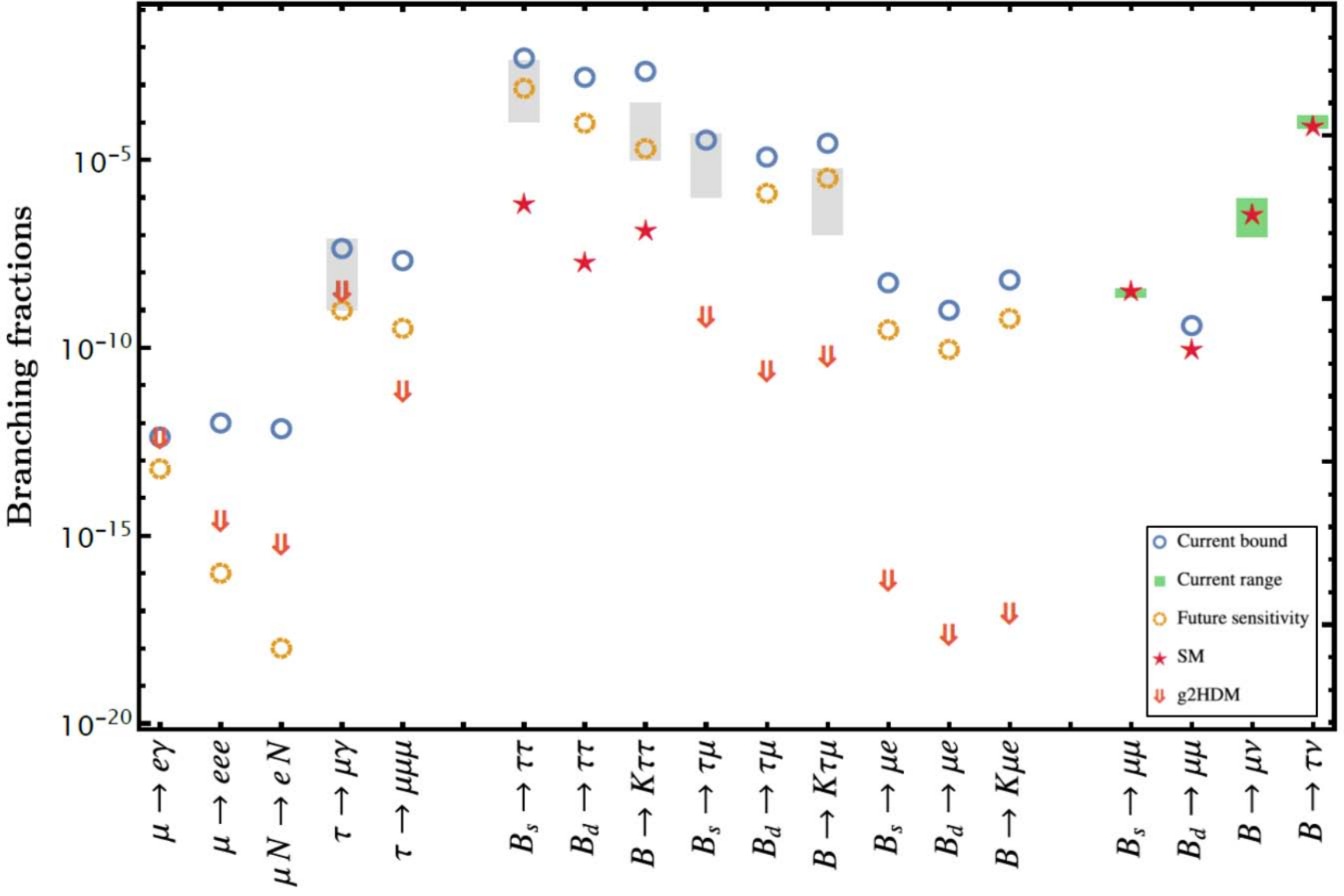}
\caption{
Pictorial table of $\mu$, $\tau$ and $B$ decay processes:
 blue solid (orange dotted) circles for current bounds
 (future sensitivities);
 green shaded bands for measured ranges of 
 $B_s \to \mu\mu$ and $B \to \tau\nu,\, \mu\nu$; 
 grey shaded bands illustrate the five 
 leading predictions of the PS$^3$ model for B-anomalies;
 red $\star$ for SM predictions.
The red $\Downarrow$ illustrate g2HDM benchmark 
projections, using 
 {$c_\gamma = 0.05$, $m_{H,\, A} = 300$~GeV, 
 $\rho_{\mu e} = \lambda_e$, 
 $\rho_{\tau\mu}=\lambda_\tau$, and 
 $\rho_{ii} = \lambda_i$}, except $\rho_{tt} = 0.4$. 
See text for more detail.}
\label{fig:contrast}
\end{figure*}

\subsection{Glimpse of the Coming New Flavor Era}

What hides $H$, $A$, $H^+$ effects so well from our view?

In this subsection we will show that 
the conservative pattern of Eq.~(\ref{eq:pattern}),
which respects the mass-mixing hierarchy 
of SM Yukawa couplings as revealed by the
$e$EDM cancellation mechanism, {\it does} hide 
exotic Higgs effects rather well in the flavor sector.

After pointing out~\cite{ref-journal30} 
the importance of $t \to ch$, we utilized 
the two-loop mechanism, which is quite analogous 
to Figs.~\ref{fig:de_BZ} and \ref{fig:de_Hgam},
to explore $\mu \to e\gamma$~\cite{ref-journal39}; 
a sizable $\rho_{tt}$, together with $\rho_{\mu e}$, 
could make it dominate over one-loop. The same 
mechanism was later applied to $\tau \to \mu\gamma$.

With myriads of extra Yukawa couplings for up- and
down-type quarks and charged leptons, we recently made 
a survey~\cite{ref-journal33} of processes of interest, 
starting with $\mu \to e\gamma$ and $\tau \to \mu\gamma$.
We chose to saturate the ranges of Eq.~(\ref{eq:pattern}).
Specifically, we take:
 $c_\gamma = 0.05$, 
 $\rho_{\mu e} = \lambda_e$, 
 $\rho_{\tau\mu} = \lambda_\tau$,
 $\rho_{ii} = \lambda_i$, except 
  $\rho_{tt} \sim 0.4$, as we take 
  the relatively low $m_{H,A} = 300$\;GeV.
We see from the left side entries of
Fig.~\ref{fig:contrast} that both 
$\mu \to e\gamma$ and $\tau \to \mu\gamma$
could be discovered in the near future,
by MEG II and Belle II, respectively.
The $\mu \to 3e$ and $\tau \to 3\mu$ decays 
would be dominated by dipole transition, with 
$\tau \to 3\mu$ falling outside of Belle II sensitivity.
Particularly interesting~\cite{ref-journal33} may be 
$\mu N \to eN$ conversion, where 
COMET at KEK and Mu2e at Fermilab aim for 
up to 6 orders of magnitude improvement. 
If realized, these experiments have the potential 
to disentangle various extra Yukawa couplings 
by utilizing different nuclei.

Turning to $B$ decays with leptons in the final state,
we contrast g2HDM with 
five spectacular projections~\cite{ref-journal40}
from leptoquarks (LQ) of the PS$^3$ model
(three copies of Pati-Salam symmetry) motivated 
by the ``B anomalies'' (for a description 
and critique, see Ref.~\cite{ref-journal41}),
which are illustrated by grey bands on the
upper side of Fig.~\ref{fig:contrast}.
The B anomalies are large effects, 
hence lead to spectacular projections,
including $\tau \to \mu\gamma$ that
falls into the Belle~II range.
We comment on the correlated modes of 
$B_s \to \tau\tau$ and $B \to K\tau\tau$ below,
where the studies are a bit more difficult.
But the two other modes,
$B_s \to \tau\mu$ (note that it was LHCb measurement
that pushed down the PS$^3$ projection!) 
and $B \to K\tau\mu$ are very interesting. 
For the latter, BaBar has shown the way with 
full hadronic tag of the other $B$, and one can just 
count events in the $m_\tau$ window, while LHCb 
has demonstrated they can do something similar,
i.e. with full kinematic control from
a decaying excited $B_s$ parent.
Surprisingly, Belle has not shown anything so far,
and $B \to K\tau\mu$ would be a competition
between LHCb and Belle II in the future.

We note from Fig.~\ref{fig:contrast} that 
$B_{s,d} \to \tau\tau$ and $B \to K\tau\tau$
have SM projections that are orders of
magnitude below experimental sensitivity,
so PS$^3$ enhancement is certainly motivating.
But for the ``middle-ground'' modes in
Fig.~\ref{fig:contrast}, g2HDM projections that are
illustrated by the red downward arrow are even 
further away from experimental scrutiny.
This illustrates the efficacy of mass-mixing 
hierarchies in hiding the exotic Higgs boson 
effects in g2HDM, making $B_{s,d} \to \mu e$ 
and $B \to K\mu e$ even harder to see,
although they certainly should be searched for.

Finally, we come to the last four modes on 
the far right of Fig.~\ref{fig:contrast}:
$B_{s,d} \to \mu\mu$ and $B^- \to \mu\nu,\,\tau\nu$.
The former two have been vigorously searched for
by LHCb and CMS, with LHCb holding the upper hand
so far, and with indication that $B_{s} \to \mu\mu$
is slightly below SM expectation, while
$B_d \to \mu\mu$ is not yet measured.
The latter two modes have been searched for 
at the B factories, with $B^- \to \tau\nu$
providing~\cite{ref-journal42} one of the 
two important bounds on $H^+$ in 2HDM~II, where 
the current result is consistent with SM expectation.
The $B^- \to \mu\nu$ mode has been under Belle
scrutiny lately~\cite{ref-journal4}, 
and will be a mode of great interest at Belle~II, 
especially in g2HDM. It was in fact 
the study~\cite{ref-journal43} of this decay that 
clarified for us some intricacies of $H^+$ effects 
in g2HDM that differs from 2HDM~II. But it still
took us some time to propose~\cite{ref-journal38} 
the $cg \to bH^+ \to bt\bar b$ process of 
Eq.~(\ref{eq:bH+}), which enjoys CKM enhancement.

We had pointed out earlier~\cite{ref-journal44}
that g2HDM could in principle make the ratio 
${\cal B}(B \to \mu\nu)/{\cal B}(B \to \tau\nu)$
deviate from the SM expectation of 0.0045,
where 2HDM~II shares the same value~\cite{ref-journal42}.
But it was only by checking explicitly~\cite{ref-journal43}
that we found that, indeed, $B \to \tau\nu$
would be SM-like, but $B \to \mu\nu$
could be more easily shifted,
even when one respects Eq.~(\ref{eq:pattern})!
Besides a large CKM enhancement on the quark side 
($b \to u$ transition), this is in part due to 
some intricacy that the {\it neutrino flavor} is not 
measured in this decay, hence 
allowing $\rho_{\tau\mu}$ to enter.
It would be exciting if Belle~II finds
${\cal B}(B \to \mu\nu)/{\cal B}(B \to \tau\nu) 
\neq 0.0045$, as it would not only be BSM,
but would rule out 2HDM~II as well.

These last four modes share the virtue that
they have SM expectations that have been driving
experimental measurement. Thus, BSM effects are
more effectively probed through interference,
and there is much to look forward to
in the near future.

\subsection{Lattice Connection:
Phase Transition and Landau Ghost}

We have mentioned that
first order electroweak phase transition
demands ${\cal O}(1)$ Higgs quartic couplings
of the g2HDM Higgs potential.
At the same time, $\mu_{22}^2/v^2$ also
ought to be ${\cal O}(1)$, otherwise a large
$\mu_{22}^2$ would damp away dynamical effects
such as EWPT, as well as EWBG itself.
Thus, $v \cong 246$\,GeV sets the electroweak scale,
and all other dimensionless parameters
of the Higgs potential are ${\cal O}(1)$,
which is why the exotic Higgs bosons populate 
300 to 600~GeV, ripe for the LHC to explore.

The ${\cal O}(1)$ Higgs quartics are not weak, 
bringing in two aspects to ponder.
The first is to go beyond one-loop 
resummed effective potential~\cite{ref-journal10}
and put the Higgs potential on 
the lattice~\cite{ref-journal45}, to check 
nonperturbatively how the first order EWPT occurs.
Though there are quite a few Higgs quartic couplings
in g2HDM, this is a question of interest 
in its own right, and it is quite timely
to check the Higgs quartic coupling parameter space 
that support first order EWPT.

A second issue is even more dynamical:
the Landau pole of these 
${\cal O}(1)$ Higgs quartics.
A simple estimate in Ref.~\cite{ref-journal12} 
gave 10--20 TeV, which is rather interesting.
It implies some strong interaction
and one would have to reconsider the theoretical
framework --- another issue to be studied on the lattice.
Establishing the strengthening of the
quartic couplings with scale would imply
New Physics at some higher scale beyond,
which could justify the 100 TeV pp collider.
Although it would be a challenge to theorists
to find a theory that can accommodate this,
one might bring back SUSY and reconsider it 
above this scale.

With enough experimental progress, one could
in principle interface the lattice study
with experimental development on exotic Higgs search,
e.g. exploring exotic Higgs scattering processes
to actually {\it measure} the 
increase in quartic coupling strength.

In any case, searching for sub-TeV
exotic Higgs bosons at the LHC is mandatory.

\begin{figure}[h]
\center
\includegraphics[width=0.27 \textwidth]{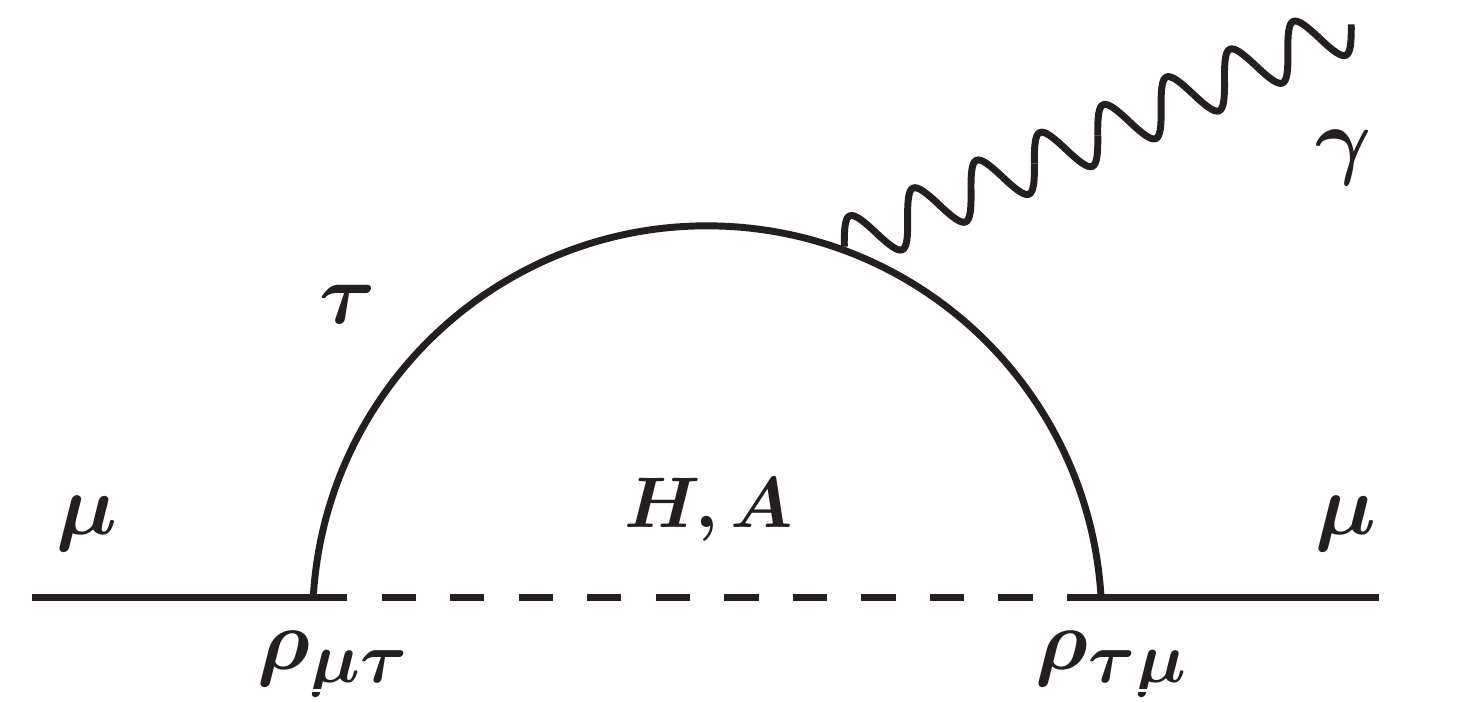}
\caption{
One-loop mechanism for muon $g-2$ anomaly,
with $\rho_{\tau\mu}$, $\rho_{\mu\tau}$
much larger than Eq.~(\ref{eq:pattern}).
}
\label{fig:1-loop}
\end{figure}

\subsection{Possible Implications of Muon g-2 in g2HDM}

After much effort, 
the Fermilab Muon g-2 experiment 
recently confirmed~\cite{ref-journal46} 
the previous measurement at BNL,
and the combined result deviates from
the ``consensus'' theory prediction~\cite{ref-journal47}
by 4.2$\sigma$. 
Thus, the muon $g-2$ anomaly has to be taken seriously.
In fact, g2HDM can quite easily handle it,
{\it if} one is willing to deviate from the 
conservative {\it guessimate} of Eq.~(\ref{eq:pattern}).
As illustrated in Fig.~\ref{fig:1-loop},
the one-loop mechanism readily handles
the observed discrepancy with, e.g.
$m_H = 300$\, GeV (with $m_A = m_{H^+}$ heavier) 
and $\rho_{\tau\mu} = \rho_{\mu\tau} \sim 0.2$,
which is 20 times $\lambda_\tau \simeq 0.01$.
The one-loop mechanism suffers chiral suppression, 
which is one of the reasons why 
the two loop mechanism could win over.
But having $\tau$ in the loop helps, as 
it is not so severely chiral-suppressed.

A large $\rho_{\tau\mu}$ or $\rho_{\mu\tau}$ 
would enter $\tau \to \mu\gamma$ via 
the two-loop mechanism with the help of $\rho_{tt}$,
analogous to our previous discussion.
But what may be surprising~\cite{ref-journal48} is that,
through the production chain~\cite{ref-journal49} of
\begin{equation}
 gg \to H,\,A \to \tau\mu,
\label{eq:ggtaumu}
\end{equation}
the recent search bounds by CMS~\cite{ref-journal50}
turns out to be more stringent on
the product $\rho_{tt}\rho_{\tau\mu}$
than from the recent Belle measurement 
of $\tau \to \mu\gamma$~\cite{ref-journal51}.
This means that a hint could well appear
with full Run~2 data, as CMS only used
36\,fb$^{-1}$ data at 13 TeV in Ref.~\cite{ref-journal50}.

With $\rho_{\tau\mu} = \rho_{\mu\tau} \sim 0.2$
and $m_H$ as before, the present bound
on $\rho_{tt}$ is about 0.1, which is 
still efficient for EWBG.
A finite $\rho_{tc}$ value would have
$t\bar c$ in the final state and dilute
the $H \to \tau\mu$ branching fraction,
hence allow $\rho_{tt}$ to be larger
(although $\tau \to \mu\gamma$ 
 constraint would kick in).
With $\rho_{tc}$ and $\rho_{\tau\mu}$
both sizable, one could have~\cite{ref-journal48} 
the novel signatures of 
$cg \to bH^+ \to \mu\tau bW^+$, $tcbW^+$ 
via $H^+ \to HW^+$ weak decay, showing 
the potential implications of the muon $g-2$ anomaly.

We found~\cite{ref-journal52} further 
profound impact of the muon $g-2$ anomaly: 
a possible revival of muon physics.
If one replaces the final $\mu$ in Fig.~\ref{fig:1-loop}
by $e$, the electron, one has one-loop mechanism
for $\mu\to e\gamma$, which would be handily suppressed
by Eq.~(\ref{eq:pattern}), but now
can allow MEG~II to probe $\rho_{\tau e}$
down to $\lambda_e$ strength.
If MEG~II makes a discovery, it can be
followed up by COMET/Mu2e for $\mu N \to eN$,
which now can even probe $\rho_{qq}$
by using various different nuclei.
For $\tau$ physics, Belle~II can readily
probe down to $\rho_{\tau\tau} = \lambda_\tau$
with $\tau \to \mu\gamma$, 
while $\tau \to 3\mu$ can now probe down to 
$\rho_{\mu\mu} \sim \lambda_\mu$,
which seems quite exciting.

We had already been investigating the EDM of
muon and tau, assuming Eq.~(\ref{eq:pattern}).
It is easy to note that the same 
one-loop diagram of Fig.~\ref{fig:1-loop}
would give rise to $d_\mu$ with 
complex $\rho_{\tau\mu}\rho_{\mu\tau}$.
We note in passing that, while this does 
not help $\tau$EDM, we find that~\cite{ref-journal34}
$\mu$EDM can be enhanced to 
$6 \times 10^{-23}\; e$\,cm, within range
of a proposed experiment at PSI,
which adds to the ``renaissance of muon physics''.

We stress, however, that 
this one-loop muon $g-2$ mechanism, 
though exciting (and not impossible), 
would actually make {\it Nature} appear 
rather ``whimsical''~\cite{ref-journal34}.
Judging from the pattern of 
hiding exotic scalar effects so well 
through the fermion mass-mixing hierarchies, 
we think that Eq.~(\ref{eq:pattern}) is
more likely to be realized.

\section{Summary}

We have presented the picture where exotic 
$H$, $A$, $H^+$ Higgs bosons exist at 500~GeV scale, 
can generate baryon asymmetry of the Universe 
while accommodating electron EDM bound 
of ACME 2018 --- and can be verified at the LHC.
This fantastic possibility is accompanied by 
a host of flavor physics and CPV probes!

We advocate, therefore:

{\begin{center}\large
 \underline{\it A Decadal Mission}
 
 {Find the extra $H$, $A$, $H^+$ bosons
 and crack the flavor code.}
 
 \normalsize Go {ATLAS/CMS} 
 \& {LHCb/Belle}\;II~(and {others})!
 
 And {Lattice}, too.
\end{center}}

We raise the following challenge: 
Having some large CPV at the {\it electroweak} scale 
to face ``the Heavens'',
how difficult is it to evade low energy
precision electron EDM probes? 
An existence proof is provided by 
the general 2HDM with extra Yukawa couplings.

\vspace{6pt}

\funding{
This research is funded by MOST 110-2639-M-002-002-ASP
of Taiwan, and also NTU 111L104019, 111L894801.}

\acknowledgments{
I have enjoyed the collaborative works with my
able collaborators as listed in the References.}

\conflictsofinterest{
The author declares no conflict of interest.
}

\abbreviations{Abbreviations}
{
The following abbreviations are used in this manuscript:\\

\begin{tabular}{@{}ll}
BSM & Beyond the Standard Model\\
NNP & No New Physics, or No New Particles\\
ALPs & Axion-like particles\\
LLPs & Long-lived particles\\
EFT & Effective Field Theory\\
CPV & CP Violation\\
EWBG & Electroweak Baryogenesis\\
EWPT & Electroweak Phase Transition\\
BAU & Baryon Asymmetry of the Universe\\
LE & Low Energy\\
HE & High Energy\\
EDM & Electric Dipole Moment\\
CKM & Cabibbo-Kobayashi-Maskawa\\
NFC & Natural Flavor Conservation\\
FCNH & Flavor Changing Neutral Higgs\\
2HDM & Two Higgs Doublet Model\\
g2HDM & general 2HDM
\end{tabular}
}

\appendixtitles{yes} 
\appendixstart
\appendix
\section[\appendixname~\thesection]
{CPV with Four Generations}

In a previous iteration, also 
based on extra Yukawa couplings, 
we advocated~\cite{ref-journal7} 
the fourth generation to provide 
sufficient CPV for baryogenesis.

There appears to be only three fermion
generations in the SM, as ``predicted''
by Kobayashi-Maskawa.
The measure of the strength of CPV
is the Jarlskog invariant~\cite{ref-journal53}, 
defined as 
${\rm Im\,det}\,[m_um_u^\dagger,\,m_dm_d^\dagger]$.
In expanded form, one has
\begin{equation}
 J = (m_t^2 - m_c^2)(m_t^2 - m_u^2)(m_c^2 - m_u^2)
     (m_b^2 - m_s^2)(m_b^2 - m_d^2)(m_s^2 - m_d^2)\,A,
\label{eq:Jarlskog}
\end{equation}
where $A$ is twice the area of 
any triangle formed by $VV^\dagger = I$,
where $V$ is the CKM matrix.

As an experimentalist constructing
the Belle detector in the 1990's, 
and also a theorist that has worked on CP violation,
the ``foklore'' is that CPV in the KM model 
falls short of what is needed for baryogenesis
by at least $10^{10}$, and it seemed that
one would prefer the B factories
to {\it reject} the KM model.
We ended up confirming it, handing
the two gentlemen their well-deserved prize.
CKM matrix of quark sector is part of SM.

It is easy to see the very strong suppression
of the Jarlskog invariant, $J$, 
in Eq.~(\ref{eq:Jarlskog}):
the two powers of $m_t^2$ are fine,
but then one accumulates the powers of
$m_b^4 m_c^2 m_s^2$, when normalized to $v^8$
gives rise to enormous suppression.
As any degeneracy within same fermion
charge would bring one back to effective
two generations, there would be no CPV.
Therefore the entry of every possible mass-squared
difference, which are really Yukawa couplings,
give rise to this severe suppression.

The fourth generation had suffered 
quite a few deaths over the years.
But another opportunity arose,
when Belle started to see a hint of
direct CPV difference between
$B^+ \to K^+\pi^0$ and $B^0 \to K^+\pi^-$,
where it could be due to the 
$Z$ penguin~\cite{ref-journal54}, which feeds 
the charge mode (by $Z$ to $\pi^0$ conversion) 
but not the neutral mode.

By 2007 or so, the situation had turned 
rather significant, and Belle was preparing
a Nature paper~\cite{ref-journal55}, 
first of its kind for B physics.
As one of the principle authors, 
as we contemplated how to present the result 
to the general reader, serendipity struck 
and we fortuitously checked the
Jarlskog invariant with four generations.
Now, the first two generations are quite close
to degenerate on the $v$-scale, being so light.
Thus, an easy way to check was to
truncate the first two generations,
which we denoted as $J_{234}$,
\begin{equation}
 J_{234} =
 (m_{t'}^2 - m_t^2)(m_{t'}^2 - m_c^2)(m_t^2 - m_c^2)
 (m_{b'}^2 - m_b^2)(m_{b'}^2 - m_s^2)(m_b^2 - m_s^2)
 \,A_{234},
\label{eq:J234}
\end{equation}
where $A_{234}$ is the truncated ``triangle''.
One sees now that only one power of 
$m_b^2$ suppression is left, and
using experimental bounds at that time,
$J_{234}$ jumped over $J$ 
by $10^{15}$~\cite{ref-journal7},
which was truly staggering.
So our basic input to a ``phenomenological''
argument of the significance of the $B^+$--$B^0$ 
direct CPV difference was to promote the 
$Z$ penguin possibility, which would be new physics, 
versus a very enhanced ``color-suppressed'' 
amplitude $C$, which would be ``hadronic effect''.

In the meantime, we had correlated~\cite{ref-journal56}
the four generation CPV effect of the $Z$ penguin
with the box diagram contribution to $B_s^0$ mixing,
more or less predicting a ``large and negative''
CPV phase in $B_s^0$--$\bar B_s^0$ oscillations.
When at the end of 2007, the CDF experiment 
at Fermilab ``saw''~\cite{ref-journal57} 
a fuzzy indication of such, the fourth generation 
gained currency~\cite{ref-journal58}.
And because of the start-up accident 
with superconducting dipole magnets, 
causing a delay of the LHC by about a year,
the Tevatron experiments pushed search limits for
fourth generation quarks to beyond the $tW$ threshold.

Alas, by 2011, the LHCb experiment did 
not~\cite{ref-journal59} find large and negative CPV 
in $B_s^0$ mixing, so $A_{234}$ is small. 
But this did not deter us, as the 
enhancement from Yukawa factors in 
Eq.~(\ref{eq:J234}) is rather large.
It was the discovery of $h(125)$ itself
(even beforehand, in limits on the cross section)
that damped the enthusiasm for the fourth generation,
because not only was the $h$ boson discovered,
its cross section matched that of
a top triangle loop in gluon-gluon fusion,
rather than the factor of 3 expected with
the presence of two additional quarks 
with masses at the weak scale.

By now, we know there has been NNP discovered 
at the LHC, other than the $h$ boson.

The moral of this recount is:
even if data-driven and experimentally motivated,
Nature may not take the path as indicated.
So, with all humility, 
although we have presented a suitably plausible
picture of an extra Higgs doublet carrying
extra Yukawa couplings that could cover the CPV
``{\bf for the Heavens and the Earth}'', 
this does not mean that it is necessarily
the path that Nature would actually follow.

\vskip0.28cm
Subtle is the Lord, but malicious He is not.

Lucidity of Eq.~(\ref{eq:simpCPV}) vs. 
Eq.~(\ref{eq:J234}) is notable. 

First order EWPT is thrown in as a bonus.

\ \ --- \;So let us finish the walk at the LHC.

\begin{adjustwidth}{-\extralength}{0cm}

\reftitle{References}

\end{adjustwidth}

\begin{thebibliography}{999}
%
\bibitem[LHCP-Reece(2021)]{ref-proceeding1} 
Reece, M. The Current State of SUSY and Ways Forward. 
LHCP2021, Paris (online), 7-12 June 2021.
%
\bibitem[Irastorza(2018)]{ref-journal1}
See e.g. Irastorza, I.G.; Redondo, J. New experimental approaches in the search for axion-like particles.
{\em Prog. Part. Nucl. Phys.} {\bf 2017}, {\em 102}, 89--159.
%
\bibitem[Curtin(2019)]{ref-journal2}
See e.g. Curtin, D. {\it et al.} Long-Lived Particles at the Energy Frontier: The MATHUSLA Physics Case.
{\em Rept. Prog. Phys.} {\bf 2019}, {\em 82}, 116201.
%
\bibitem[Arcadi(2018)]{ref-journal3}
See e.g. Arcadi, G. 
{\it et al.} The waning of the WIMP? A review of models, searches, and constraints.
{\em Eur. Phys. J. C} {\bf 2018}, {\em 78}, 203.
%
\bibitem[LHCP-Pomarol(2021)]{ref-proceeding2}
Pomarol, A. The SM EFT \& new physics. 
LHCP2021, Paris (online), 7-12 June 2021.
%
\bibitem[ParticleDataGroup(2020)]{ref-journal4}
 Zyla, P.A. {\it et al.} Review of Particle Physics.
{\em PTEP} {\bf 2020}, {\em 2020}, 083C01.
%
\bibitem[Sakharov(1967)]{ref-journal5}
Sakharov, A. Violation of CP Invariance, C asymmetry, and baryon asymmetry of the universe.
{\em Pisma Zh. Eksp. Teor. Fiz.} {\bf 1967}, {\em 5}, 32--35.
%
\bibitem[Andreev(2018)]{ref-journal6}
Andreev, V. {\it et al.} [ACME] Improved limit on the electric dipole moment of the electron.
{\em Nature} {\bf 2018}, {\em 7727}, 355--360.
%
\bibitem[Hou(2009)]{ref-journal7}
Hou, W.-S. Source of CP Violation for the Baryon Asymmetry of the Universe.
{\em Chin. J. Phys.} {\bf 2009}, {\em 47}, 134--141.
%
\bibitem[Glashow(1977]{ref-journal8}
Glashow, S.L.; Weinberg, S. Natural Conservation Laws for Neutral Currents.
{\em Phys. Rev. D} {\bf 1977}, {\em 15}, 1958--1965.
%
\bibitem[Aad(2016)]{ref-journal9}
Aad, G. \textit{et al.} [ATLAS and CMS] Measurements of the Higgs boson production and decay rates and constraints on its couplings from a combined ATLAS and CMS analysis of the LHC pp collision data at $\sqrt{s}=7$ and 8 TeV.
{\em JHEP} {\bf 2016}, {\em 08}, 045.
%
\bibitem[Tumasyan(2021)]{ref-unpublish1}
Tumasyan, A. {\it et al.} [CMS] Search for flavor-changing neutral current interactions of the top quark and Higgs boson in final states with two photons in proton-proton collisions at $\sqrt{s}$ = 13 TeV. 
\textit{arXiv:2111.02219 [hep-ex]}, \textit{submitted to Phys. Rev. Lett.}.
%
\bibitem[Kanemura(2005)]{ref-journal10}
See e.g. Kanemura, S.; ~Okada, Y.; Senaha, E. Electroweak baryogenesis and quantum corrections to the triple Higgs boson coupling.
{\em Phys. Lett. B} {\bf 2005}, {\em 606}, 361--366.
%
\bibitem[Davidson(2005)]{ref-journal11}
Davidson, S.; Haber, H.E. Basis-independent methods for the two-Higgs-doublet model.
{\em Phys. Rev. D} {\bf 2005}, {\em 72}, 035004.
%
\bibitem[Hou(2018)]{ref-journal12}
Hou, W.-S.; Kikuchi, M. Approximate Alignment in Two Higgs Doublet Model with Extra Yukawa Couplings.
{\em EPL} {\bf 2018}, {\em 123}, 11001.
%
\bibitem[Fuyuto(2018)]{ref-journal13}
Fuyuto, K.; Hou, W.-S.; Senaha, E. Electroweak baryogenesis driven by extra top Yukawa couplings.
{\em Phys. Lett. B} {\bf 2018}, {\em 776}, 402--406.
%
\bibitem[Chiang(2016)]{ref-journal14}
Chiang, C.W.; Fuyuto, K.; Senaha, E. Electroweak Baryogenesis with Lepton Flavor Violation.
{\em Phys. Lett. B} {\bf 2016}, {\em 762}, 315--320.
%
\bibitem[Ade(2014)]{ref-journal15}
Ade, P.A.R. \textit{et al.} [Planck] Planck 2013 results. XVI. Cosmological parameters.
{\em Astron. Astrophys.} {\bf 2014}, {\em 571}, A16.
%
\bibitem[Guo(2017)]{ref-journal16}
Guo, H.-K.; Li, Y.-Y.; Liu, T.; Ramsey-Musolf, M.; Shu, J.  Lepton-Flavored Electroweak Baryogenesis.
{\em Phys. Rev. D} {\bf 2017}, {\em 96}, 115034.
%
\bibitem[Altunkaynak(2015)]{ref-journal17}
Altunkaynak, B.; Hou, W.-S., Kao, C.; Kohda M.; McCoy, B.  Flavor Changing Heavy Higgs Interactions at the LHC.
{\em Phys. Lett. B} {\bf 2015}, {\em 751}, 135--142.
%
\bibitem[Barr(1990]{ref-journal18}
Barr, S.M.; Zee, A. Electric Dipole Moment of the Electron and of the Neutron.
{\em Phys. Rev. Lett.} {\bf 1990}, {\em 65}, 21--24.
%
\bibitem[Baron(2014)]{ref-journal19}
Baron, J. \textit{et al.} [ACME] Order of Magnitude Smaller Limit on the Electric Dipole Moment of the Electron.
{\em Science} {\bf 2014}, {\em 343}, 269--272.
%
\bibitem[Baron(2017)]{ref-journal20}
Baron, J. \textit{et al.} [ACME] Methods, Analysis, and the Treatment of Systematic Errors for the Electron Electric Dipole Moment Search in Thorium Monoxide.
{\em New J. Phys.} {\bf 2017}, {\em 19}, 073029.
%
\bibitem[Fuyuto(2018)]{ref-journal21}
Fuyuto, K.; Hou, W.-S.; Senaha, E. Cancellation mechanism for the electron electric dipole moment connected with the baryon asymmetry of the Universe.
{\em Phys. Rev. D} {\bf 2020}, {\em 101}, 011901(R).
%
\bibitem[Abe(2014)]{ref-journal22}
Abe, T.; Hisano, J.; Kitahara, T.; Tobioka, K. Gauge invariant Barr-Zee type contributions to fermionic EDMs in the two-Higgs doublet models.
{\em JHEP} {\bf 2014}, {\em 01}, 106.
%
\bibitem[Chupp(2019)]{ref-journal23}
Chupp, T.; Fierlinger, P.; Ramsey-Musolf, M.; Singh, J. Electric dipole moments of atoms, molecules, nuclei, and particles.
{\em Rev. Mod. Phys.} {\bf 2019}, {\em 91}, 015001.
%
\bibitem[Fuyuto(2019)]{ref-journal24}
Fuyuto, K.; Ramsey-Musolf, M.; Shen, T. Electric Dipole Moments from CP-Violating Scalar Leptoquark Interactions.
{\em Phys. Lett. B} {\bf 2019}, {\em 788}, 52--57.
%
\bibitem[Dekens(2019)]{ref-journal25}
Dekens, W.; de Vries, J.; Jung, M.; Vos, K.K. The phenomenology of electric dipole moments in models of scalar leptoquarks.
{\em JHEP} {\bf 2019}, {\em 01}, 069.
%
\bibitem[Cesarotti(2019)]{ref-journal26}
Cesarotti, C.; Lu, Q.; Nakai, Y.; Parikh, A.; Reece, M. Interpreting the Electron EDM Constraint.
{\em JHEP} {\bf 2019}, {\em 05}, 059.
%
\bibitem[Albrecht(1987)]{ref-journal27}
Albrecht, H. \textit{et al.} [ARGUS] Observation of $B^0$-$\bar B^0$ Mixing.
{\em Phys. Lett. B} {\bf 1987}, {\em 192}, 245--252.
%
\bibitem[Wolfenstein(1983)]{ref-journal28}
Wolfenstein, L. Parametrization of the Kobayashi-Maskawa Matrix.
{\em Phys. Rev. Lett.} {\bf 1983}, {\em 51}, 1945--1947.
%
\bibitem[Cheng(1987)]{ref-journal29}
Cheng, T.P.; Sher, M. Mass Matrix Ansatz and Flavor Nonconservation in Models with Multiple Higgs Doublets.
{\em Phys. Rev. D} {\bf 1987}, {\em 35}, 3484--3491.
%
\bibitem[Hou(1992)]{ref-journal30}
Hou, W.-S. Tree level $t \to c h$ or $h \to t \bar c$ decays.
{\em Phys. Lett. B} {\bf 1992}, {\em 296}, 179--184.
%
\bibitem[Cairncross(2017)]{ref-journal31}
Cairncross, W.B.; Gresh, D.N.; Grau, M.; Cossel, K.C.; Roussy, T.S.; Ni, Y.; Zhou, Y.; Ye, J.; Cornell, E.A.  Precision Measurement of the Electron's Electric Dipole Moment Using Trapped Molecular Ions.
{\em Phys. Rev. Lett.} {\bf 2017}, {\em 119}, 153001.
%
\bibitem[Bean(1987)]{ref-journal32}
Bean, A. \textit{et al.} [CLEO] Limits on $B^0$-$\bar B^0$ Mixing and $\tau_{B^0}/\tau_{B^+}$.
{\em Phys. Rev. Lett.} {\bf 1987}, {\em 58}, 183--186.
%
\bibitem[Hou(2020)]{ref-journal33}
Hou, W.-S.; Kumar, G. Muon Flavor Violation in Two Higgs Doublet Model with Extra Yukawa Couplings.
{\em Phys. Rev. D} {\bf 2020}, {\em 102}, 115017.
%
\bibitem[Hou(2022)]{ref-journal34}
Hou, W.-S.; Kumar, G.; Teunissen, S. Charged Lepton EDM with Extra Yukawa Couplings.
{\em JHEP} {\bf 2022}, {\em 01}, 092.
%
\bibitem[Kohda(2018)]{ref-journal35}
Kohda, M.; Modak, T.; Hou, W.-S. Searching for new scalar bosons via triple-top signature in $cg \to tS^0 \to tt\bar t$.
{\em Phys. Lett. B} {\bf 2018}, {\em 776}, 379--384.
%
\bibitem[Hou(2021)]{ref-journal36}
Hou, W.-S.;  Modak, T. Probing Top Changing Neutral Higgs Couplings at Colliders.
{\em Mod. Phys. Lett. A} {\bf 2021}, {\em 36}, 21300064.
%
\bibitem[Hou(2019)]{ref-journal37}
Hou, W.-S.; Kohda, M.; Modak, T. Implications of Four-Top and Top-Pair Studies on Triple-Top Production.
{\em Phys. Lett. B} {\bf 2019}, {\em 798}, 134953.
%
\bibitem[Ghosh(2020)]{ref-journal38}
Ghosh, D.K.; Hou, W.-S.;  Modak, T. Sub-TeV $H^+$ Boson Production as Probe of Extra Top Yukawa Couplings.
{\em Phys. Rev. Lett.} {\bf 2020}, {\em 1256}, 221801.
%
\bibitem[Chang(1993)]{ref-journal39}
Chang, D.; Hou, W.-S.; Keung, W.-Y. Two loop contributions of flavor changing neutral Higgs bosons to $\mu \to e \gamma$.
{\em Phys. Rev. D} {\bf 1993}, {\em 48}, 217--224 
%
\bibitem[Bordone(2018)]{ref-journal40}
Bordone, M.; Cornella, C.; Fuentes-Mart\' in, J.; Isidori, G. Low-energy signatures of the $\mathrm{PS}^3$ model: from $B$-physics anomalies to LFV.
{\em JHEP} {\bf 2018}, {\em 10}, 148.
%
\bibitem[Hou(2019)]{ref-journal41}
Hou, G.W.-S. Perspectives and Outlook from HEP Window on the Universe.
{\em Int. J. Mod. Phys. A} {\bf 2019}, {\em 34}, 1930002.
%
\bibitem[Hou(1993)]{ref-journal42}
Hou, W.-S. Enhanced charged Higgs boson effects in $B \to \tau \bar\nu,\, \mu \bar\nu$ and $b \to \tau \bar\nu + X$.
{\em Phys. Rev. D} {\bf 1993}, {\em 48}, 2342--2344.
%
\bibitem[Hou(2020)]{ref-journal43}
Hou, W.-S.; Kohda, M.; Modak, T.; Wong, G.-G. Enhanced $B \to \mu \bar\nu$ decay at tree level as probe of extra Yukawa couplings.
{\em Phys. Lett. B} {\bf 2020}, {\em 800}, 135105.
%
\bibitem[Chang(2017)]{ref-journal44}
Chang, P.; Chen, K.-F.; Hou, W.-S. Flavor Physics and CP Violation.
{\em Prog. Part. Nucl. Phys.} {\bf 2017}, {\em 97}, 261--311.
%
\bibitem[Kainulainen(2019)]{ref-journal45}
Kainulainen, K.; Keus, V.; Niemi, L.; Rummukainen, K.; Tenkanen, T.V.I.; Vaskonen V. On the validity of perturbative studies of the electroweak phase transition in the Two Higgs Doublet model.
{\em JHEP} {\bf 2019}, {\em 06}, 075.
%
\bibitem[Abi(2021)]{ref-journal46}
Abi, B. \textit{et al.} [Muon g-2] Measurement of the Positive Muon Anomalous Magnetic Moment to 0.46 ppm.
{\em Phys. Rev. Lett.} {\bf 2021}, {\em 126}, 141801.
%
\bibitem[Aoyama(2020)]{ref-journal47}
Aoyama, T. \textit{et al.} The anomalous magnetic moment of the muon in the Standard Model.
{\em Phys. Rept.} {\bf 2020}, {\em 887}, 1--166.
%
\bibitem[Hou(2021)]{ref-journal48}
Hou, W.-S.; Jain, R.; Kao, C.; Kumar, G.; Modak, T. Collider Prospects for Muon $g-2$ in General Two Higgs Doublet Model.
{\em Phys. Rev. D} {\bf 2021}, {\em 104}, 075036.
%
\bibitem[Hou(2019)]{ref-journal49}
Hou, W.-S.; Jain, R.; Kao, C.; Kohda, M.; McCoy, B.; Soni, A. Flavor Changing Heavy Higgs Interactions with Leptons at Hadron Colliders.
{\em Phys. Lett. B} {\bf 2019}, {\em 795}, 371--378.
%
\bibitem[Sirunyan(2020)]{ref-journal50}
Sirunyan, A.M. \textit{et al.} [CMS] Search for lepton flavour violating decays of a neutral heavy Higgs boson to $\mu\tau$ and e$\tau$ in proton-proton collisions at $\sqrt{s}=$ 13 TeV.
{\em JHEP} {\bf 2020}, {\em 03}, 103.
%
\bibitem[Abdesselam(2021)]{ref-journal51}
Abdesselam, A. \textit{et al.} [Belle] Search for lepton-flavor-violating tau-lepton decays to $\ell\gamma$ at Belle.
{\em JHEP} {\bf 2021}, {\em 10}, 19.
%
\bibitem[Hou(2021)]{ref-journal52}
Hou, W.-S.; Kumar, G. Charged lepton flavor violation in light of muon $g-2$.
{\em Eur. Phys. J. C} {\bf 2021}, {\em 81}, 1132.
%
\bibitem[Jarlskog(1985)]{ref-journal53}
Jarlskog, C. Commutator of the Quark Mass Matrices in the Standard Electroweak Model and a Measure of Maximal CP Violation.
{\em Phys. Rev. Lett.} {\bf 1985}, {\em 55}, 1039--1042.
%
\bibitem[Hou(2005)]{ref-journal54}
Hou, W.-S.; Nagashima, M.; Soddu, A. Difference in $B^+$ and $B^0$ direct CP asymmetry as effect of a fourth generation.
{\em Phys. Rev. Lett.} {\bf 2005}, {\em 95}, 141601.
%
\bibitem[Lin(2008)]{ref-journal55}
Lin, S.-W.; Unno, Y.; Hou, W.-S.; Chang P. \textit{et al.} [Belle] Difference in direct charge-parity violation between charged and neutral B meson decays.
{\em Nature} {\bf 2008}, {\em 452}, 332--335.
%
\bibitem[Hou(2007)]{ref-journal56}
Hou, W.-S.; Nagashima, M.; Soddu, A. Large time-dependent CP violation in $B_s^0$ system and finite $D^0$--$\bar{D}^0$ mass difference in four generation standard model.
{\em Phys. Rev. D} {\bf 2007}, {\em 76}, 016004.
%
\bibitem[Aaltonen(2008)]{ref-journal57}
Aaltonen, T. \textit{et al.} [CDF] First Flavor-Tagged Determination of Bounds on Mixing-Induced CP Violation in $B^0_{s} \to J/\psi \phi$ Decays.
{\em Phys. Rev. Lett.} {\bf 2008}, {\em 100}, 161802.
%
\bibitem[Holdom(2009)]{ref-journal58}
Holdom, B.; Hou, W.-S.; Hurth, T.; Mangano, M.L.;   Sultansoy, S.; \" Unel, G. Four Statements about the Fourth Generation.
{\em PMC Phys. A} {\bf 2009}, {\em 3}, 4.
%
\bibitem[Aaij(2012)]{ref-journal59}
Aaij, R. \textit{et al.} [LHCb] Measurement of the CP-violating phase $\phi_s$ in the decay $B^0_s\to J/\psi \phi$.
{\em Phys. Rev. Lett.} {\bf 2012}, {\em 108}, 101803.
%

\end{thebibliography}
\end{document}